 \documentclass[10pt]{article}

\usepackage{amssymb}
\usepackage{lineno}

\usepackage{amsmath}
\usepackage{subfigure}
\usepackage{color}

\usepackage{graphics} % for pdf, bitmapped graphics files
\usepackage{epsfig} % for postscript graphics files
\usepackage{mathptmx} % assumes new font selection scheme installed
\usepackage{times} % assumes new font selection scheme installed
\usepackage{amsmath} % assumes amsmath package installed
\usepackage{amssymb}  % assumes amsmath package installed
\usepackage{graphicx}
\usepackage[ansinew]{inputenc}
\usepackage{longtable}
\usepackage{subfigure}
\usepackage{algorithm,algorithmic}
\usepackage{epsfig}
\usepackage{float}
\usepackage{color}
\usepackage{dsfont}
\usepackage{bm}

\usepackage{calrsfs}

\usepackage{xfrac}
\usepackage{tikz}

\DeclareMathAlphabet{\pazocal}{OMS}{zplm}{m}{n}

\newcommand{\dif}{\textrm{d}}
\DeclareMathOperator{\erf}{erf}

\DeclareMathOperator{\imag}{Im}

\DeclareMathOperator{\real}{Re}
\DeclareMathOperator{\cov}{cov} \DeclareMathOperator{\ex}{E}

\title{ A Three Spatial Dimension Wave Latent Force Model for Describing Excitation Sources and Electric Potentials Produced by Deep Brain Stimulation.}

\usepackage{authblk}
\author[1]{Pablo A. Alvarado \thanks{This manuscript presents in more detail the research on LFM for DBS carried out the period during which Pablo A. Alvarado pursued MSc studies at Universidad Tecnol\'ogica de Pereira, Colombia. Preliminary results were publised at The IEEE Engineering in Medicine and Biology Conference EMBC 2014, (see \cite{Alvarado14}).}}
\author[2]{Mauricio A. \'Alvarez}
\author[2]{\'{A}lvaro A. Orozco}
\affil[1]{Queen Mary University of London, UK}
\affil[2]{Universidad Tecnol\'ogica de Pereira, Colombia}

\date{}

\begin{document}
\maketitle
%\begin{frontmatter}



%\address[label1]{Queen Mary University of London}
%\address[label2]{Universidad Tecnol\'{o}gica de Pereira}

%\author{}
%\address{}

\begin{abstract}

\noindent
Deep brain stimulation (DBS) is a surgical treatment for Parkinson's Disease. Static models based on quasi-static approximation are common approaches for DBS modeling. While this simplification has been validated for bioelectric sources, its application to rapid stimulation pulses, which contain more high-frequency power, may not be appropriate, as DBS therapeutic results depend on stimulus parameters such as frequency and pulse width, which are related to time variations of the electric field.
We propose an alternative hybrid approach based on probabilistic models and differential equations, by using Gaussian processes and wave equation. Our model avoids quasi-static approximation, moreover, it is able to describe dynamic behavior of DBS. Therefore, the proposed
model may be used to obtain a more realistic phenomenon description. The proposed model can also solve inverse problems, i.e. to recover the corresponding source of excitation, given electric potential distribution.
The electric potential produced by a time-varying source was predicted using proposed model. For static sources, the electric potential produced by different electrode configurations were modeled. Four different sources of excitation were recovered by solving the inverse problem.
We compare our outcomes with the electric potential obtained by solving Poisson's equation using the Finite Element Method (FEM).
Our approach is able to take into account time variations of the source and the produced field. Also, inverse problem can be addressed using the proposed model. The electric potential calculated with the proposed model is close to the potential obtained by solving Poisson's equation using FEM.

%\noindent
%Deep brain stimulation is a surgical treatment for Parkinson's Disease. Its fundamental purpose is to deliver electrical stimulation in a target brain structure using implanted electrodes.
%
%Static models based on the quasi-static approximation are the most common approach used for deep brain stimulation modeling.  While this simplification has been validated for bioelectric sources, its application to rapid stimulation pulses, which contain more high-frequency power, may not be appropriate, as therapeutic results of deep brain stimulation are quite dependent on stimulus parameters such as frequency and pulse width, which are related to time variations of the electric field.  In this work we propose an alternative hybrid approach based on probabilistic models and partial differential equations, by using Gaussian processes and the wave equation. Our model avoids the quasi-static approximation, moreover, it is able to describe dynamic behavior inherent to deep brain stimulation. Therefore, the proposed model may be used to obtain a more realistic phenomenon description. 
%
%Furthermore, the proposed model can be used for solving the inverse problem, i.e. given the electric potential distribution, it is possible to recover the corresponding source of excitation, which is a valuable clinical application, as it would allow appropriate tuning of the DBS device by the expert physician.
\end{abstract}

%\begin{keyword}

%Deep brain stimulation \sep Gaussian processes \sep latent force model \sep wave equation.

%\end{keyword}

%\end{frontmatter}

%\linenumbers

\section{Introduction}

\noindent Parkinson's disease (PD) is a degenerative disorder of the central nervous system. Its effects are defective motor skills and speech. PD is the second most common neurodegenerative disorder after Alzheimer's disease, most frequently affecting elderly population. The treatment for PD includes medication, physical therapy, and surgical procedures such as deep brain stimulation (DBS) \cite{Chaovalitwongse11}. 
DBS is the preferred surgical treatment for symptoms of advanced PD when they are no longer controlled with just drug
therapy \cite{Davidson14,Benabid09}. The purpose of DBS is to deliver electrical stimulation in a specific brain structure, using implanted electrodes % connected to a stimulator
%, through which a train of constant voltage or current square pulses are applied to stimulate the surrounding neural tissue  
\cite{Benabid03,Yousif10}.
%With this aim, in a stereotactic surgery, a stimulation electrode is implanted in a group of nuclei which are responsible for pathological effects of PD.
The common nuclei used for treatment are the subthalamic nucleus (STN), and globus pallidus pars interna (GPi), which are situated at the base of the forebrain \cite{Schmidt12}. 

DBS can also result in significant declines in the cognitive and cognitive-motor performance of PD patients, because of the spread of current to non-motor areas of the STN or adjacent brain structures.
One of the most common causes of unsuccessful DBS therapy is an inadequately parameters configuration \cite{McIntyre09}.
%
%Suitable parameters programming depends on several factors, including maximizing therapeutic efficacy, minimizing side effects and prolonging battery life. 
%
Although guidelines exist on typically effective DBS frequencies, pulse widths, and the most common electrode configurations, the
variability among patients limits the use of this information \cite{Montgomery10}.
Also, it is not practical to clinically
evaluate each of the thousands of possible stimulation
parameter combinations. 
That is why simulation using computational models of the electric propagation induced by DBS appears as an useful 3D visualization tool for assisting the clinical programming process \cite{Schmidt12,McIntyre09}.

Electric fields generated by DBS are dynamic, since a real DBS stimulus corresponds to an square waveform with a fundamental frequency range from 130 Hz up to 185 Hz  \cite{Kuncel04,Walckiers10,Volkmann02,Osuilleabhain03}. Nevertheless, electric potential  induced close to the stimulating electrode is commonly modeled using Laplace \cite{Schmidt12,Butson11,Liberti07}, or Poisson \cite{Miocinovic09,Butson05,Walckiers10} equation, assuming a quasi-static or static field. The quasi-static approximation neglects wave propagation effects and time derivatives in Maxwell's equations, simplifying the models by avoiding time variations \cite{Bossetti08}. In these models, the source is represented as static and its dynamic behavior is discarded. 
This simplification has been validated for bioelectric
sources, but it may not be appropriate for stimulation pulses with high-frequency components \cite{Bossetti08}.

A Fourier Finite Element Method (Fourier FEM) that takes into account dynamics in DBS was presented
in \cite{Butson05}.
Despite the fact that the approach implemented in \cite{Butson05} takes into account the time, Fourier FEM gives steady state solutions
and does not model transients, that is, effects of wave propagation are neglected.
Furthermore,
in \cite{Bossetti08} the authors compared potentials calculated using the quasi-static approximation (Poisson's equation) with those calculated from the inhomogeneous Helmholtz wave equation in an infinite, homogeneous, and isotropic volume conductor using a point current source stimulus. In \cite{Bossetti08}  the implemented methodology uses the time variable, but its results were obtained assuming an infinite domain.  

In this work we introduce a novel latent force model (LFM) \cite{Alvarez09} based on the wave equation in three spatial dimensions. This LFM allows to describe time variations of both, the source as well as the electric potential produced by DBS.
A LFM is a strongly mechanistic non-parametric probabilistic model, that combines Gaussian processes (GPs) with differential equations in a machine learning approach \cite{Alvarez09}.
%
%hybrid approach based on probabilistic models and partial differential equations, by using latent force models using Gaussian processes \cite{Alvarez09}.
%
The main goal is to solve a partial differential equation (PDE)
subject to some boundary constraints by using
GPs \cite{Alvarez13}. 
%
%In this case, the GPs represent random variables that correspond first to excitation source values at any time and location within the solution domain, and second to values of the electric potential at any place of the tissue medium at any time \cite{Rasmussen05}. 
In particular, we are solving the second order nonhomogeneous wave equation with three space variables in the rectangular Cartesian system of coordinates, in a  rectangular
parallelepiped domain \cite{Polyanin02}. 
%Our proposal is a general formulation of the electric propagation problem. In fact, by restricting our model it would be possible to obtain the Poisson's formulation. 
%

The main advantage  of the proposed model is that offers an alternative approach that can deal with the calculation of the electric potential produced by DBS, taking into account propagation effects and time-varying sources of excitation,  in a three-dimensional finite domain. 
In comparison with the FEM approximated solution, which does not have a close form, i.e. the computed FEM solution is only valid for specific system parameters values and a given source of excitation, 
in the proposed wave LFM we are formulating a general expression for calculating the probability distribution over the wave equation solution conditioned to the source of excitation.
This allows to use the model for predicting the electric potential produced by different sources. Also, with the proposed approach we can study how the conditional distribution over the equation solution is affected by changes in the system parameters, as well as modifications in the covariance function hyperparameters associated with the latent force stochastic process.
%
%
%
%
%
%
%Furthermore, in the proposed LFM the input stimulus is not limited to be a point source.
%
Finally, our approach is able to solve the inverse problem \cite{Sarkka11}, i.e.
given the electric potential distribution, it is possible to recover
the corresponding input stimulus and its parameters, which
is a valuable clinical application, as it would allow appropriate
tuning of the DBS device by the expert physician.

The organization of this paper is as follows: in Section 2 we
introduce the theory of some electromagnetic models widely used for describing the electric potential produced during DBS. Then, we present the proposed latent force model and the formulation of its covariance and cross-covariance functions. In section 3 we present results obtained from different simulations, by using a forward problem as well as an inverse problem approach. Finally, in
	section 4 the conclusions are presented.
\section{Materials and methods}

\noindent 
Most approaches for electric potential modelling in DBS are based on either the Laplace or the Poisson equation \cite{Schmidt12}  \cite{Walckiers10}
 \cite{Liberti07} \cite{Miocinovic09} \cite{Butson05}
  \cite{Grant09}. This requires to assume the electric potential field as quasi-static \cite{Bossetti08}. The quasi-static approximation simplifies the
wave equation for the electric potential by neglecting the second partial derivative with respect to time \cite{Steinmetz09}
%used to describe the wave propagation, defined as %
%
\begin{align}\label{e:wave}
\nabla^2f  - \frac{1}{c^2} \frac{\partial^2 f }{\partial t^2 } = -\frac{\rho}{\varepsilon},
	\end{align}
where $f$ is the electric potential, $c$ is the propagation
velocity of the electromagnetic wave, $\rho$ denotes the
electric space charge density, and $\varepsilon$ is the
permittivity \cite{Hermann89}. Therefore, the wave equation reduces to the Poisson equation \cite{Bossetti08} \cite{Sadiku02},
\begin{align}\label{e:poisson}
\nabla^2f  = -\frac{\rho}{\varepsilon}.
\end{align}
Furthermore, if we consider no sources  we get the Laplace equation
\begin{align}\label{e:laplace}
\nabla^2f   = 0.
\end{align}
From \eqref{e:poisson} and \eqref{e:laplace} it is evident that the quasi-static approximation limits the models to not take
into account time variations \cite{Bossetti08}.
%
%\subsection{Wave Latent Force Model for describing DBS}
%
In order to avoid the quasi-static approximation, we present a novel LFM based on the wave equation for describing the electric potential produced during DBS. We use the general expression for the second order inhomogeneous wave equation with three space variables in the rectangular Cartesian system of coordinates.

In the next sections we introduce general concepts of LFM using GPs, then we provide theory about the second order non-homogeneous wave equation, and its solution using Green's functions. Finally, we present the mathematical formulation of the covariance and cross-covariance functions of the proposed LFM.

\subsection{Gaussian Processes and Latent Force Models}
\noindent
Gaussian processes (GPs) are probability distributions over functions, such as any finite set of function evaluations (i.e. a collection of random variables) follows a jointly Gaussian distribution \cite{Rasmussen05}. 
The underlying idea in LFMs is to combine a
physically-inspired model together with a probabilistic distribution over latent functions called forces
\cite{Alvarez13}. Specifically, the LFM presented here uses the wave equation as mechanistic model, and the forces represent the DBS source of excitation.
In a LFM, GPs
are used to describe two functions, i.e. the source of excitation
$u(\textbf{x},t)$ as well as the solution $f(\textbf{x},t)$ of the differential equation implemented, in this case the electric potential at location $\textbf{x}$ at time $t$, where  $\textbf{x}\in\mathbb{R}^D$, with $D=2$ or $3$ in rectangular Cartesian coordinates.
Specifically, we use GPs for defining a
probabilistic prior over the latent function $u(\textbf{x},t)$ .
The latent force $u(\textbf{x},t)$ follows a GP prior, assuming zero mean and kernel
$k_{u}(\textbf{x},\textbf{x}';t,t')$, that is
\begin{align}\label{e:prior_u}
u(\textbf{x},t) \sim \pazocal{GP}(0, k_{u}(\textbf{x},\textbf{x}';t,t')).
\end{align}
The wave differential equation is a linear operator. Therefore the result of applying this operator to the latent force, i.e. the solution $f(\textbf{x},t)$ of the wave equation, also corresponds to a GP with zero mean and covariance function $k_{f}(\textbf{x},\textbf{x};t,t')$, that is
\begin{align}\label{e:prior_f}
f(\textbf{x},t) \sim \pazocal{GP}(0, k_{f}(\textbf{x},\textbf{x}';t,t')).
\end{align}
The cross-covariance function $k_{fu}(\textbf{x},\textbf{x}';t,t')$
between $f(\textbf{x},t)$ and $u(\textbf{x},t)$ is also calculated. Details about the calculation of these covariance functions can be found in the appendix. 
Assuming we observe the source of excitation or latent function $u(\textbf{x},t)$ at specific times and points in space, $\boldsymbol{u} = \{ u(\textbf{x}_i,t_j), i =1:N_{\textbf{x}}, j=1:N_t \}$, where $N_{\textbf{x}}$ and $N_t$ correspond to the number of space points and time instants respectively. For simplicity let us assume we want to predict the solution of the differential equation in the same time instants and points in space, i.e. $ \boldsymbol{f} = \{ f(\textbf{x}_i,t_j), i =1:N_{\textbf{x}}, j=1:N_t \}$. By definition of the GP \cite{Rasmussen05}, the joint distribution of   $\boldsymbol{u}$ and $\boldsymbol{f}$ has the following form
\begin{align}\label{e:jointDis}
\begin{bmatrix}
\boldsymbol{u} \\ \boldsymbol{f}
\end{bmatrix}
\sim
\pazocal{N} 
\left(
\begin{bmatrix}
0 \\ 0
\end{bmatrix}
,
\begin{bmatrix}
K_{u} & K_{fu}^{\top}  \\  K_{fu} & K_{f}
\end{bmatrix}
\right),
\end{align}
where the matrices $K_{u}$, $K_{f}$  are computed using the covariance function defined for the prior over the latent force i.e. $k_{u}(\textbf{x},\textbf{x}';t,t')$, and the kernel obtained from the solution of the partial differential equation i.e. $k_{f}(\textbf{x},\textbf{x}';t,t')$ respectively. Matrix $K_{fu}$ is calculated using the cross-covariance function $k_{fu}(\textbf{x},\textbf{x}';t,t')$ mentioned before. Using the properties for multivariate Gaussian distributions \cite{Bishop06}, we can get the posterior distribution over the solution of the differential equation given an specific source of excitation. Also we can compute the posterior distribution over the latent function given a prescribed solution of the differential equation.
If we want to  get the conditional distribution over the
collection of random variables $\boldsymbol{f}$, given an specific source of excitation
$\boldsymbol{u}$, this is known as the forward problem, and the distribution of the wave equation solution conditioned to the latent force is given by \cite{Rasmussen05},
\begin{align}\label{e:f_posterior}
  p (\boldsymbol{f}|\boldsymbol{u}) \sim \pazocal{N} \left( K_{fu} K_{u}^{-1} \boldsymbol{u}
\ , \
K_{f} - K_{fu}K_{u}^{-1}K_{fu}^{\top}  \right).
\end{align}
%where $K_{f}$, $K_{fu}$, and $K_{u}$ are covariance matrices computed from functions $k_{f,f}(\cdot,\cdot)$, $k_{f,u}(\cdot,\cdot)$, and $k_{u,u}(\cdot,\cdot)$, at particular space points and time instants.
%
The proposed LFM can address the inverse problem as well by conditioning the distribution over the latent force to a specific solution of the wave equation. This conditional distribution is given by
\begin{align}\label{e:u_posterior}
p (\boldsymbol{u}|\boldsymbol{f}) \sim \pazocal{N} \left( K_{fu}^{\top} K_{f}^{-1} \boldsymbol{f}
\ , \
K_{u} - K_{fu}^{\top}K_{f}^{-1}K_{fu}  \right).
\end{align}
The next section gives details about the form of the kernel functions $k_u(\textbf{x},\textbf{x}';t,t')$, $k_f(\textbf{x},\textbf{x}';t,t')$ and $k_{fu}(\textbf{x},\textbf{x}';t,t')$, which are used for computing the matrices $K_{u}$, $K_{f}$ and $K_{fu}$ in \eqref{e:f_posterior} and \eqref{e:u_posterior}.

\subsection{LFMs Using the Wave Equation}
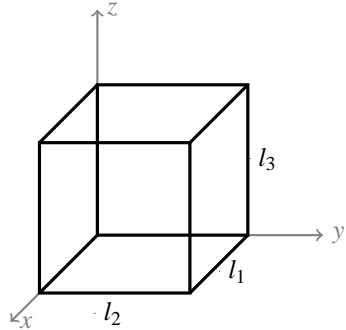
\begin{figure}[h!]
\centering
\begin{tikzpicture}
[cube/.style={very thick,black},
grid/.style={very thin,gray},
axis/.style={->,gray,thick}]
%
%draw a grid in the x-y plane
%\foreach \x in {-0.5,0,...,2.5}
%\foreach \y in {-0.5,0,...,2.5}
%{
%	\draw[grid] (\x,-0.5) -- (\x,2.5);
%	\draw[grid] (-0.5,\y) -- (2.5,\y);
%}
%
%draw the axes
\draw[axis] (0,0,0) -- (3,0,0) node[anchor=west]{$y$};
\draw[axis] (0,0,0) -- (0,3,0) node[anchor=west]{$z$};
\draw[axis] (0,0,0) -- (0,0,3) node[anchor=west]{$x$};
%
%draw the top and bottom of the cube
\draw[cube] (0,0,0) -- (0,2,0) -- (2,2,0) -- (2,0,0) -- cycle;
\draw[cube] (0,0,2) -- (0,2,2) -- (2,2,2) -- (2,0,2) -- cycle;
%
%draw the edges of the cube
\draw[cube] (0,0,0) -- (0,0,2);
\draw[cube] (0,2,0) -- (0,2,2);
\draw[cube] (2,0,0) -- (2,0,2);
\draw[cube] (2,2,0) -- (2,2,2);
\draw (1,0,2.7) -- (1,0,2.7) node[midway,right] {$l_2$};
\draw (2.1,0,1.25) -- (2.1,0,1.25) node[midway,right] {$l_1$};
\draw (2.5,1.5,1.25) -- (2.5,1.5,1.25) node[midway,right] {$l_3$};
\end{tikzpicture}
\caption{Domain of solution for the wave equation.}
\label{f:cube_domain}
\end{figure}

\noindent 
The general expression of the second order non-homogeneous wave equation \cite{Polyanin02} with three space variables in the rectangular Cartesian system of coordinates has the form 
\begin{align} \label{e:waveGeneral}
\frac{\partial^2 f}{\partial t^2} = a^2 \left( \frac{\partial^2 f}{\partial x^2}  + \frac{\partial^2 f}{\partial y^2} + \frac{\partial^2 f}{\partial z^2} \right) + S u , 
\end{align}
where $f = f(\textbf{x},t)$ is the unknown function, $a$ is a constant coefficient related with the propagation
velocity of the electromagnetic wave, $u = u(\textbf{x},t)$ is a source defined as a latent force, $S$ quantifies the influence of the latent force $u(\textbf{x},t)$ over the output $f(\textbf{x},t)$, and $\textbf{x}=[x,y,z]$.
The exact solution of \eqref{e:waveGeneral} is subject to the domain of solution, as well as particular initial and boundary conditions. For a boundary value problem with a rectangular parallelepiped as domain (see Fig. \ref{f:cube_domain}), and assuming homogeneous boundary conditions, %i.e. 
%
%\begin{align*} 
%\begin{array}{lll}
%f = f_{0}(x,y,z)  = 0                     &   \text{at } t = 0               &   \text{(initial condition)}, \\
%\partial_{t}f = f_{1}(x,y,z) = 0    &  \text{at } t = 0                &  \text{(initial condition)}, \\
%f = g_{1}(y,z,t)        =0              &  \text{at } x = 0               &    \text{(boundary condition)}, \\
%f = g_{2}(y,z,t)          =0             &   \text{at } x = l_{1}       &   \text{(boundary condition)}, \\
%f = g_{3}(x,z,t)        =0                &   \text{at } y = 0              &   \text{(boundary condition)}, \\
%f = g_{4}(x,z,t)         =0               &   \text{at } y = l_{2}       &   \text{(boundary condition)}, \\
%f = g_{5}(x,y,t)        =0              &   \text{at } z = 0              &   \text{(boundary condition)}, \\
%f = g_{6}(x,y,t)      =0                &   \text{at } z = l_{3}       &   \text{(boundary condition)}, 
%\end{array}
%\end{align*}
%
the solution to \eqref{e:waveGeneral} is given by 
\begin{align}\label{e:SolGreen}
f(\textbf{x},t)  = \int\limits_{0}^{t} \int\limits_{\boldsymbol{\rho}} u(\boldsymbol{\rho}, \tau)   
G(\textbf{x}, \boldsymbol{\rho}, t- \tau)   \dif \boldsymbol{\rho}  \dif\tau,
\end{align}
where  $\boldsymbol{\rho} = [\xi, \eta,\zeta]$, the term $S$ in \eqref{e:waveGeneral} was included inside $u$, and $G(\textbf{x},\boldsymbol{\rho},t)$  is the Green's function for the wave equation, defined as \cite{Polyanin02}
%
%\begin{align}\label{e:Green}
%\frac{8}{a l_1 l_2 l_3} 
%\sum_{n=1}^{\infty} 
%\sum\limits_{m=1}^{\infty}
%\sum\limits_{k=1}^{\infty}
%\frac{1}{\lambda_{nmk}}
%\sin(\alpha_n x)
%\sin(\beta_m y)
%\sin(\gamma_k z)
%\sin(\alpha_n \xi)
%\sin(\beta_m\eta)
%\sin(\gamma_k \zeta)
%\sin(a\lambda_{nmk}t), 
%\end{align}
%
%
\begin{align}\label{e:Green}
%G(\textbf{x},\boldsymbol{\rho},t)
%=
\sum_{n=1}^{\infty} 
\sum\limits_{m=1}^{\infty}
\sum\limits_{k=1}^{\infty}
\hat{\lambda}_{nmk}
\
g_{nmk}(\textbf{x})
\
h_{nmk}(\boldsymbol{\rho})
\
\sin(a\lambda_{nmk}t), 
\end{align}
where 
\begin{align*}
 \hat{\lambda}_{nmk} &= \frac{8}{ a l_1 l_2 l_3 \lambda_{nmk}},
\\
g_{nmk}(\textbf{x}) &= \sin(\alpha_n x)
\sin(\beta_m y)
\sin(\gamma_k z),
\\
h_{nmk}(\boldsymbol{\rho}) &= \sin(\alpha_n \xi)
\sin(\beta_m\eta)
\sin(\gamma_k \zeta),
\end{align*}
and finally \cite{Polyanin02} $\alpha_n = \sfrac{n \pi}{l_1}$, $\beta_m = \sfrac{m \pi}{l_2}$, $\gamma_k = \sfrac{k \pi}{l_3}$, $\lambda_{nmk} = \sqrt{\alpha_n^2 + \beta_m^2 + \gamma_k^2}$ .
%\begin{align*}
%\alpha_n = \frac{n \pi}{l_1}, 
%\qquad
%\beta_m = \frac{m \pi}{l_2},
%\qquad 
%\gamma_k = \frac{k \pi}{l_3},
%\qquad 
%\lambda_{nmk} = \sqrt{\alpha_n^2 + \beta_m^2 + \gamma_k^2}.
%\end{align*}
%
%Assuming homogeneous boundary conditions, i.e. $g_s \equiv 0$ ($s=1,2,3,4,5,6$),  $f_0 = 0$ and $f_1 = 0$, expression \eqref{e:w_all} is equal to
%
We assume that the source or latent function $u(\textbf{x},t)$  in \eqref{e:waveGeneral} follows a Gaussian process prior with zero mean and covariance function $k_{u}(\textbf{x},\textbf{x}';t,t')$, that is
$
u(\textbf{x},t) \sim  \pazocal{GP} (0, k_u(\textbf{x},\textbf{x}';t,t') ),
$
where the covariance function is defined as 
\begin{align}\label{e:CovSource}
k_u(\textbf{x},\textbf{x}';t,t') 
= 
%\cov[u(\textbf{x},t),u(\textbf{x}',t')]
%=
%\ex[u(\textbf{x},t)u(\textbf{x}',t')] 
%= 
k(x, x')k(y, y')k(z, z')k(t, t').
\end{align}
The kernel $k(\cdot,\cdot)$ in \eqref{e:CovSource} is prescribed to follow a squared exponential form:
\begin{align}\label{e:SquaredExp}
k(x,x') = 
\exp  \left( - \frac{(x-x')^2}{\sigma^2_x}\right),
\end{align}
where $\sigma^2_x$ is known as the length-scale  \cite{Rasmussen05}. Since the wave equation \eqref{e:waveGeneral} is linear, its solution also follows a Gaussian process. We assume that the solution to the wave equation or output $f(\textbf{x},t)$ follows a Gaussian process prior with zero mean and covariance function $k_f(\textbf{x},\textbf{x}';t,t')$, that is, 
$
f(\textbf{x},t) \sim  \pazocal{GP} (0, k_f(\textbf{x},\textbf{x}';t,t') ),
$
where the covariance function %$k_f(\textbf{x},\textbf{x}';t,t') $ 
is defined as 
\begin{align}
\label{e:CovOutput}
k_f(\textbf{x},\textbf{x}';t,t') 
= 
\cov[f(\textbf{x},t),f(\textbf{x}',t')]
=
\ex[f(\textbf{x},t)f(\textbf{x}',t')].
\end{align}
%$k_f(\textbf{x},\textbf{x}';t,t')$
%A detailed definition of the covariance function \eqref{e:CovOutput} can be found in appendix \ref{appendix}. 
%
Using (\ref{e:SolGreen}), then \eqref{e:CovOutput} can be expressed as
%
%\begin{align}
%\notag
%\ex \left[  
%\int\limits_{0}^{t} \int\limits_{0}^{l_3} \int\limits_{0}^{l_2}  \int\limits_{0}^{l_1} Su(\xi, \eta, \zeta, \tau)   
%G(\textbf{x}, \xi, \eta,\zeta, t- \tau)   \dif \xi \ \dif \eta  \ \dif\zeta \ \dif\tau
%%
%\int\limits_{0}^{t'} \int\limits_{0}^{l_3} \int\limits_{0}^{l_2}  \int\limits_{0}^{l_1} S'u(\xi', \eta', \zeta', \tau')   
%G(\textbf{x}', \xi', \eta',\zeta', t'- \tau')   \dif \xi' \ \dif \eta'  \ \dif \zeta' \ \dif \tau'
%\right],
%\end{align}
%
%there is only uncertainty over the latent function $u$, then
%
%\begin{align}
%\notag
%\int\limits_{0}^{t}  
%\int\limits_{0}^{t'} 
%\int\limits_{\boldsymbol{\rho}}  
%\int\limits_{\boldsymbol{\rho}'} 
%%\int\limits_{0}^{l_3}  
%%\int\limits_{0}^{l_3}
%%\int\limits_{0}^{l_2}  
%%\int\limits_{0}^{l_2}
%%\int\limits_{0}^{l_1} 
%%\int\limits_{0}^{l_1}  
%S^{2}
%G(\textbf{x}, \boldsymbol{\rho}, t- \tau)
%G(\textbf{x}', \boldsymbol{\rho}', t'- \tau')
%\ex [ u( \boldsymbol{\rho}, \tau)  u( \boldsymbol{\rho}', \tau')]
%\dif \boldsymbol{\rho}\
%\dif \boldsymbol{\rho}' \
%%\dif \xi'  \ \dif \xi \
%%\dif\eta'  \ \dif\eta  \ 
%%\dif\zeta' \ \dif\zeta \ 
%\dif\tau'  \ \dif\tau ,
%\end{align}
%
%
%
\begin{align}\label{e:kf}
%\int\limits_{0}^{t}  
%\int\limits_{0}^{t'} 
\int\limits_{ \boldsymbol{\hat{\tau}}}^{}
\int\limits_{\boldsymbol{\rho}}  
\int\limits_{\boldsymbol{\rho}'} 
%\int\limits_{0}^{l_3}  
%\int\limits_{0}^{l_3}
%\int\limits_{0}^{l_2}  
%\int\limits_{0}^{l_2}
%\int\limits_{0}^{l_1} 
%\int\limits_{0}^{l_1}  
%S^{2}
G(\textbf{x}, \boldsymbol{\rho}, t- \tau)
G(\textbf{x}',\boldsymbol{\rho}', t'- \tau')
\hat{k}(\boldsymbol{\rho},\boldsymbol{\rho}';\tau,\tau')
%k(\tau, \tau')
\dif\boldsymbol{\rho}\
\dif\boldsymbol{\rho}' \
%\dif \xi'   \dif \xi \
%\dif\eta'   \dif\eta  \ 
%\dif\zeta' \dif\zeta \ 
\dif \boldsymbol{\hat{\tau}}
%\dif\tau'\dif\tau,
\end{align}
where $\boldsymbol{\hat{\tau}} = [\tau, \tau']$, $\boldsymbol{\rho}' = [\xi', \eta',\zeta']$, and
\begin{align*}
\hat{k}(\boldsymbol{\rho},\boldsymbol{\rho}';\tau,\tau') = S^{2}
k(\xi, \xi')
k(\eta, \eta')
k(\zeta, \zeta')
k(\tau, \tau').
\end{align*} 
%
%Using  \eqref{e:CovSource} for the expected value in the last expression we have
%
%
The covariance function for the solution of the wave equation \eqref{e:waveGeneral} is given by the solution of \eqref{e:kf}.
The cross covariance function $k_{fu}(\textbf{x},\textbf{x}';t,t')$ between the output $f(\textbf{x},t)$ and the latent function $u(\textbf{x}',t')$% needed for the computation of the matrix $K_{fu}$ in \eqref{e:jointDis}, 
, is given by solving
%\begin{align}
%k_{fu}(\textbf{x},\textbf{x}';t,t') = 
%\cov\left[f(\textbf{x},t),u(\textbf{x}',t')\right]=
%S
%\ex \left[
%\int\limits_{0}^{t}
%\int\limits_{0}^{l_3}
%\int\limits_{0}^{l_2}
%\int\limits_{0}^{l_1}
%u(\xi, \eta, \zeta, \tau)
%u(\textbf{x}',t')
%G(\textbf{x}, \xi,\eta,\zeta, t- \tau)  \dif \xi  \  \dif \eta \ \dif \zeta \ \dif \tau 
%\right]
%\end{align}
%Then, the covariance $\cov\left[f(\textbf{x},t),u(\textbf{x}',t')\right]$ is given as
\begin{align*} 
\int\limits_{0}^{t}
\int\limits_{\boldsymbol{\rho}}
%\int\limits_{0}^{l_3}
%\int\limits_{0}^{l_2}
%\int\limits_{0}^{l_1}
G(\textbf{x},\boldsymbol{\rho}, t- \tau)
\ex \left[
u(\boldsymbol{\rho}, \tau) u(\textbf{x}',t')   
\right]
\dif \boldsymbol{\rho} \
%\dif \xi  \ \dif \eta  \ \dif \zeta \
\dif \tau.
\end{align*}
%
%where $\boldsymbol{\rho} = [\xi,\eta,\zeta]$. 
Using the factorized form for the covariance of the latent function \eqref{e:CovSource}, the last expression can be written as
\begin{align}\label{e:cross_cov}
%S 
\int\limits_{0}^{t}
\int\limits_{\boldsymbol{\rho}}
%\int\limits_{0}^{l_3}
%\int\limits_{0}^{l_2}
%\int\limits_{0}^{l_1}
G(\textbf{x},\boldsymbol{\rho},t- \tau)
k(\xi,x') k(\eta,y') k(\zeta,z') k(\tau,t')
% \dif \xi \ \dif \eta  \ \dif \zeta \ 
\dif \boldsymbol{\rho} \
\dif \tau.
\end{align}
The solution for the covariance function \eqref{e:kf} of the output of the wave equation  can be obtained analytically, as well as the solution for the cross-covariance function \eqref{e:cross_cov} between the input and the output of the wave equation. A detailed solution for both covariance functions \eqref{e:CovOutput} and \eqref{e:cross_cov} can be found in the appendix.

\section{Results and discussion}
\noindent
In this section, we present results obtained by simulating different experiments, using the proposed latent force model based on the wave equation. 
Firstly, in a forward problem approach \eqref{e:f_posterior}, we simulate the electric potential produced during DBS for different electrode configurations. 
Here the proposed LFM is compared with the Finite Element Method (FEM) solution of the Poisson equation. 
Then, in a two spatial dimension domain ($\textbf{x} \in \mathbb{R}^{2}$) the wave LFM is used for solving the inverse problem \eqref{e:u_posterior}, i.e. to compute the distribution over the DBS excitation conditioned to a prescribed electric potential.
Finally, in order to highlight that our approach is able to describe time varying fields, we show experiments in $\textbf{x} \in \mathbb{R}^{3}$ where the latent force (DBS excitation) evolves periodically in time.

\subsection{Forward Problem Approach}
\begin{figure}[h!]
	\centering
	\includegraphics[width=0.33\columnwidth]{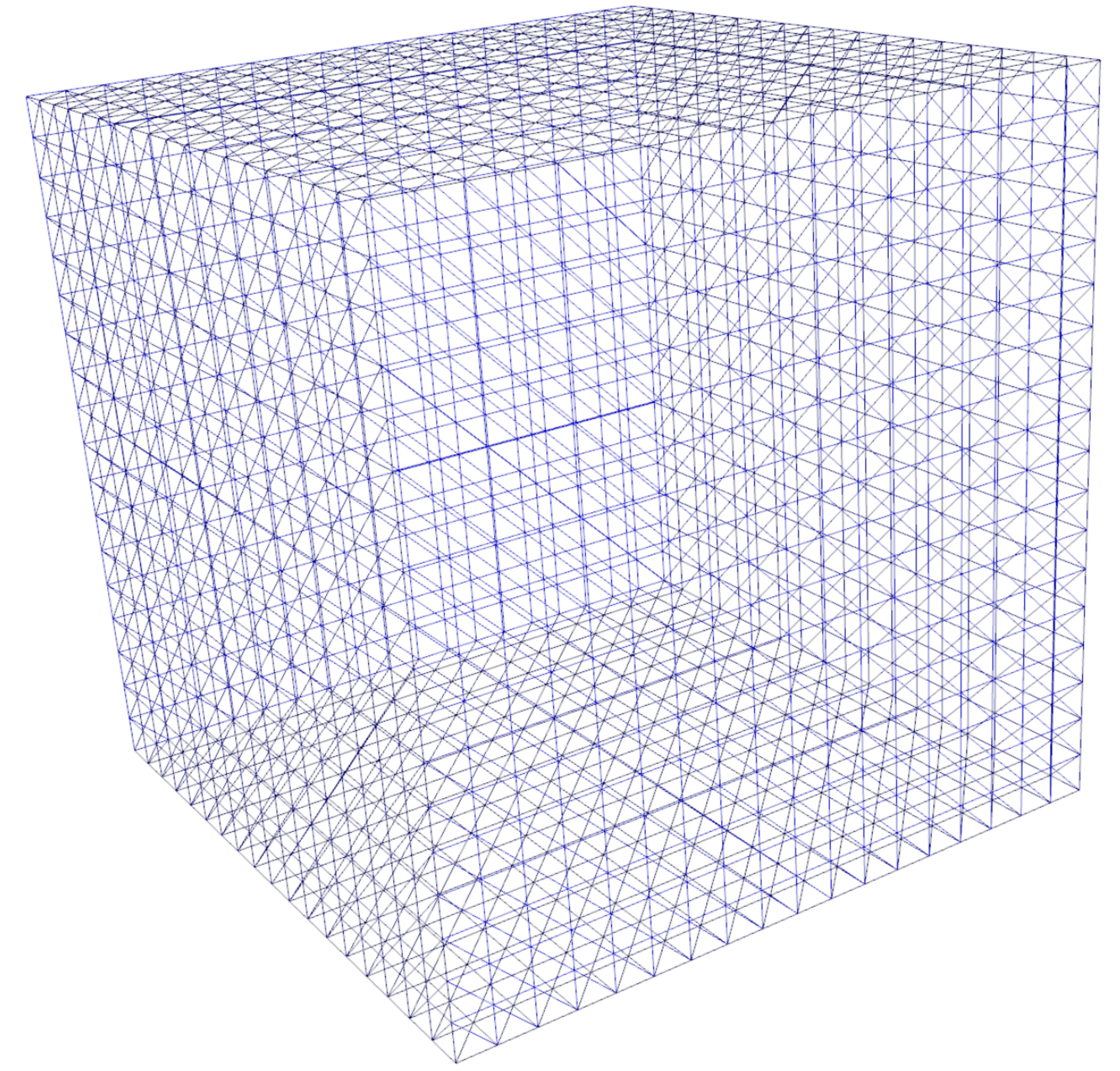}
	\caption{Solution Domain used for the forward problem.}
	\label{fig:SolutionDomain}
\end{figure}

\noindent
The domain of solution for these simulations was: an uniform mesh of $19 \times 19 \times 19$ points in a cubic domain with size $10 \text{cm} \times 10 \text{cm}  \times 10 \text{cm}$ (see Fig.\ref{fig:SolutionDomain}). This is because we were interested in simulating the electric potential within the region of interest (ROI) around the stimulation centre \cite{Schmidt12}. All boundary and initial conditions were set to zero.
Results were compared with the solution of Poisson equation using the FEM toolbox FEniCS \cite{Logg12}. 
The hyperparameters of the proposed LFM were tuned manually,
where $\sigma_{x} = \sigma_{y} =\sigma_{z} =\sigma_{t} = 0.01$ in \eqref{e:CovSource}, and $a = 1 \times 10^{5}$ in \eqref{e:waveGeneral}.
To face the drawback of $\text{O}(N^3)$ computational complexity for the prediction using GPs, and taking into account the proposed LFM kernel is a \textit{tensor product kernel} (TPK) \cite{Luo13}, we used methods presented in  \cite{Saatci11,Xu12,Luo13} to make computational savings.

\subsubsection{Simulation of Deep Brain Stimulation}\label{DirectProblem}

\begin{figure}[!h]  
	\centering
	\subfigure {\def\svgwidth{80pt} 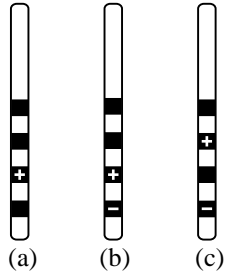}
	\caption{Examples of electrode configuration. (a) Monopolar, single contact. (b) Bipolar, single positive. (c) Bipolar, single positive \cite{Montgomery10}.}
	\label{fig:DBSexamples}
\end{figure}

\noindent
The simulations in this section were done assuming  static excitations in \eqref{e:wave}, that is  $u(\textbf{x},t) =  u(\textbf{x})$.
We simulated three commonly used electrode configurations as point sources.  Fig. \ref{fig:DBSexamples} shows the monopolar (Fig. \ref{fig:DBSexamples}(a)) and bipolar ( Fig.  \ref{fig:DBSexamples}(b)-(c)) configurations used.
Each electrode configuration was modeled as a piecewise function, defined as 
%
%$u(x,y,t) = 1 \times 10^{-3}$ 
%$u(x,y,z) =\pm 1 $ 
%
%in electrode contact locations, and $u(x,y,z) = 0$ elsewhere,  
%
%
\begin{align*}
u(x,y,z) = \left\{
\begin{array}{lr}
\pm 1 & \text{in electrode contact locations,} \\
0 & \text{elsewhere},
\end{array}
\right.
\end{align*} 
depending on which source in Fig. \ref{fig:DBSexamples} is used.

\begin{figure}[!h]
	\centering
	%\includegraphics[width=1\columnwidth]{model.eps}\\
	\subfigure[]{\includegraphics[width=0.33\columnwidth]{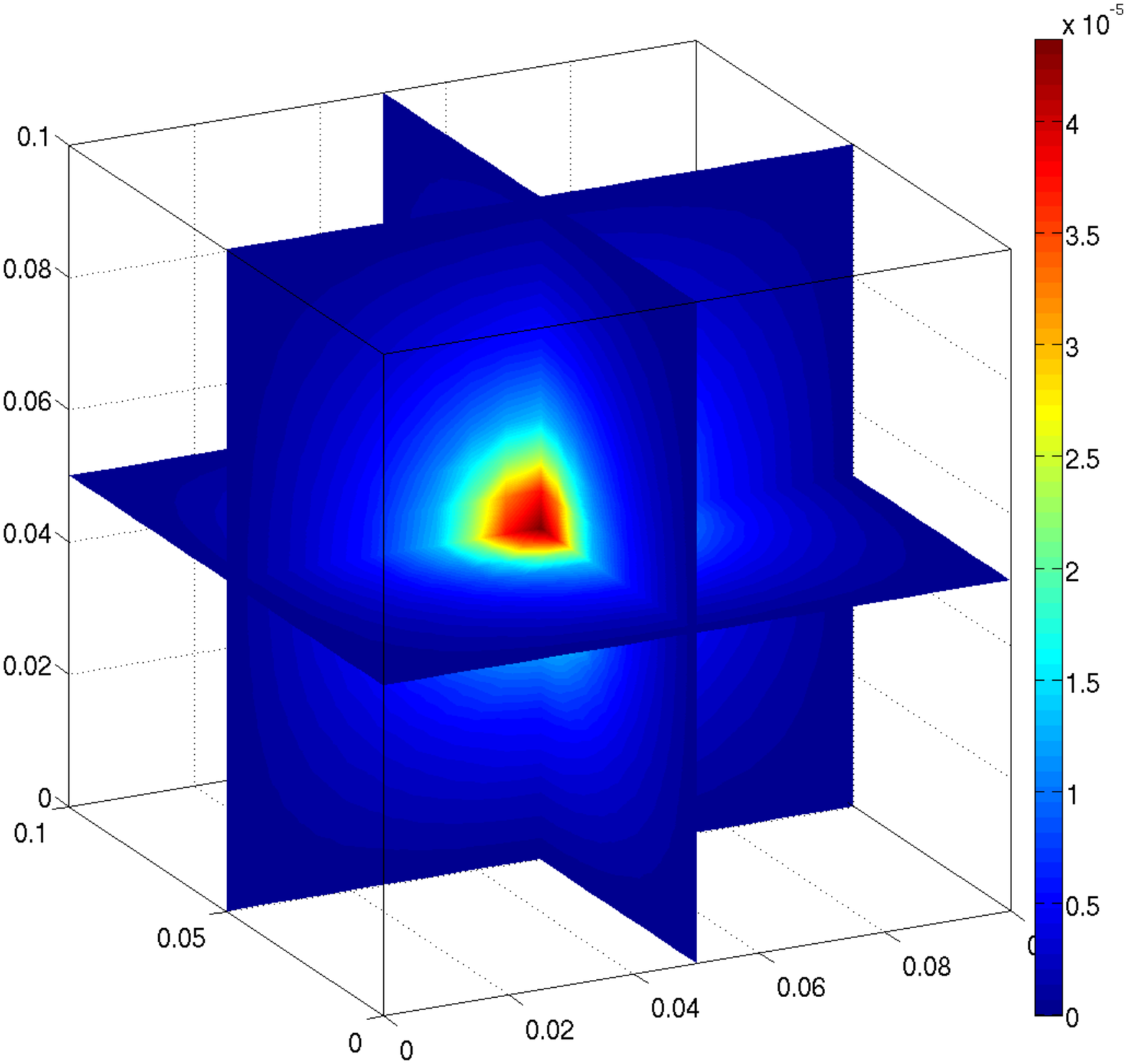}\label{fig:LFMslicesS1}}%\\
	\subfigure[]{\includegraphics[width=0.33\columnwidth]{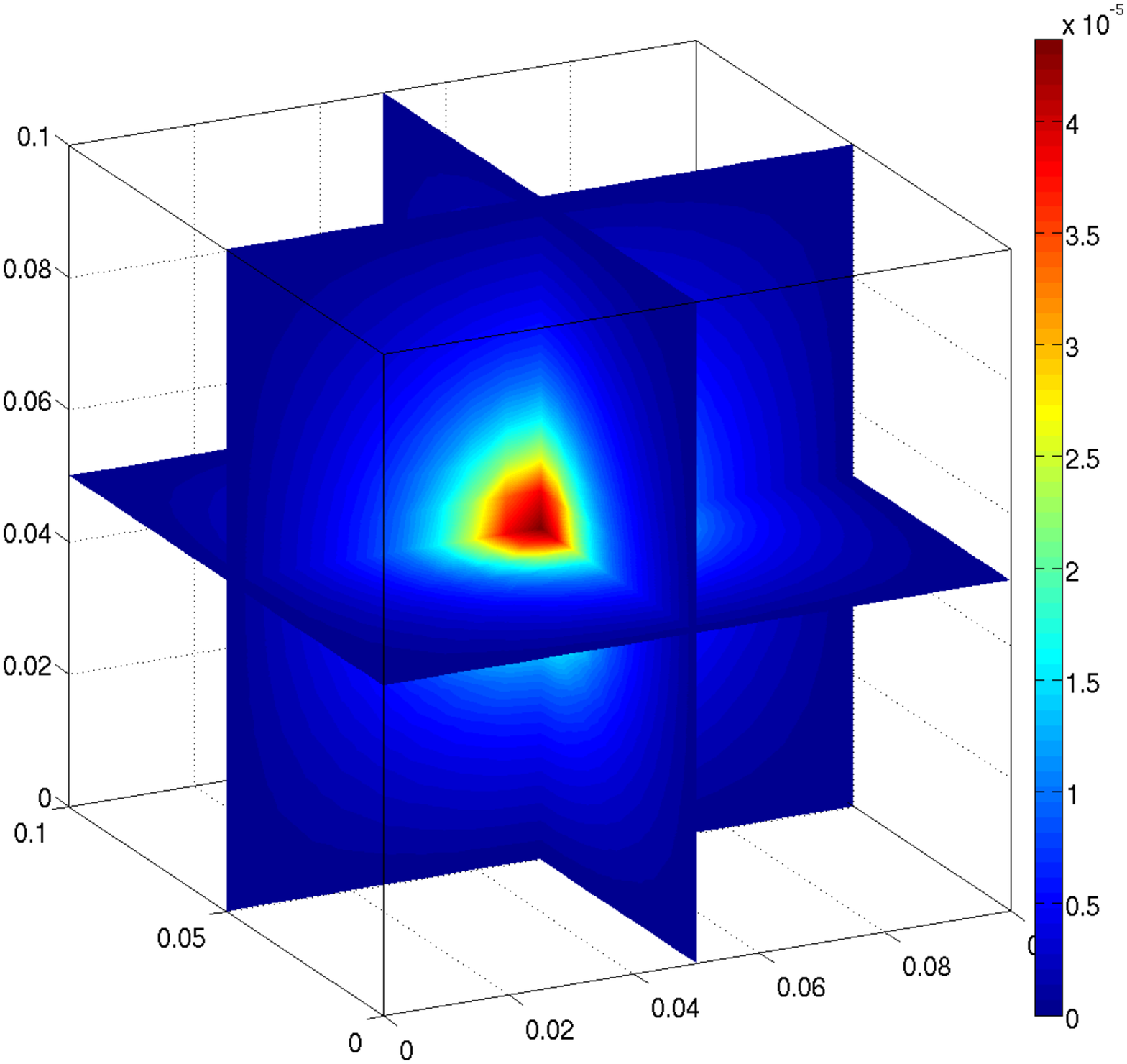}\label{fig:FEMslicesS1}}\\
	\subfigure[]{\includegraphics[width=0.33\columnwidth]{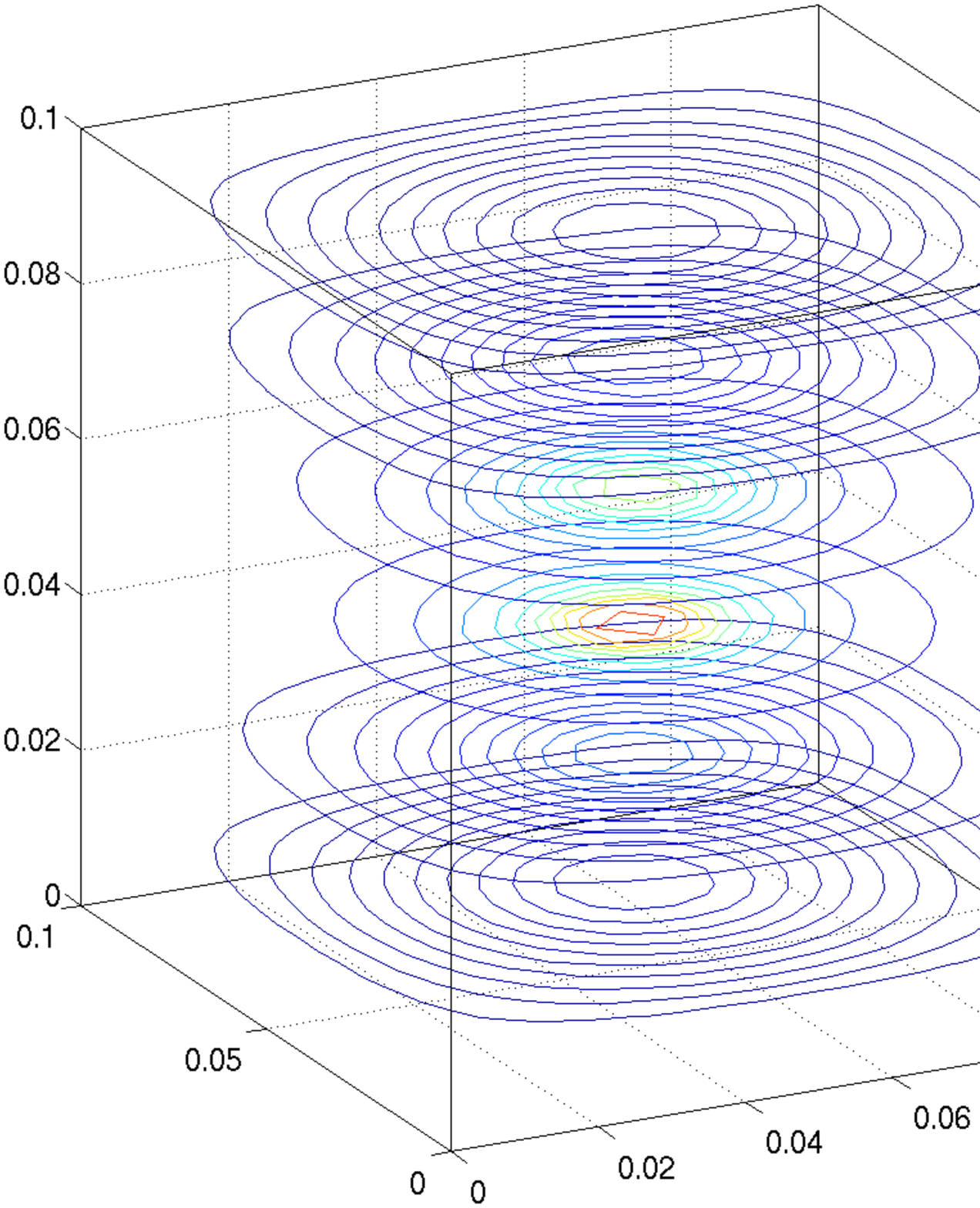}\label{fig:LFMcontourS1}}%\\
	\subfigure[]{\includegraphics[width=0.33\columnwidth]{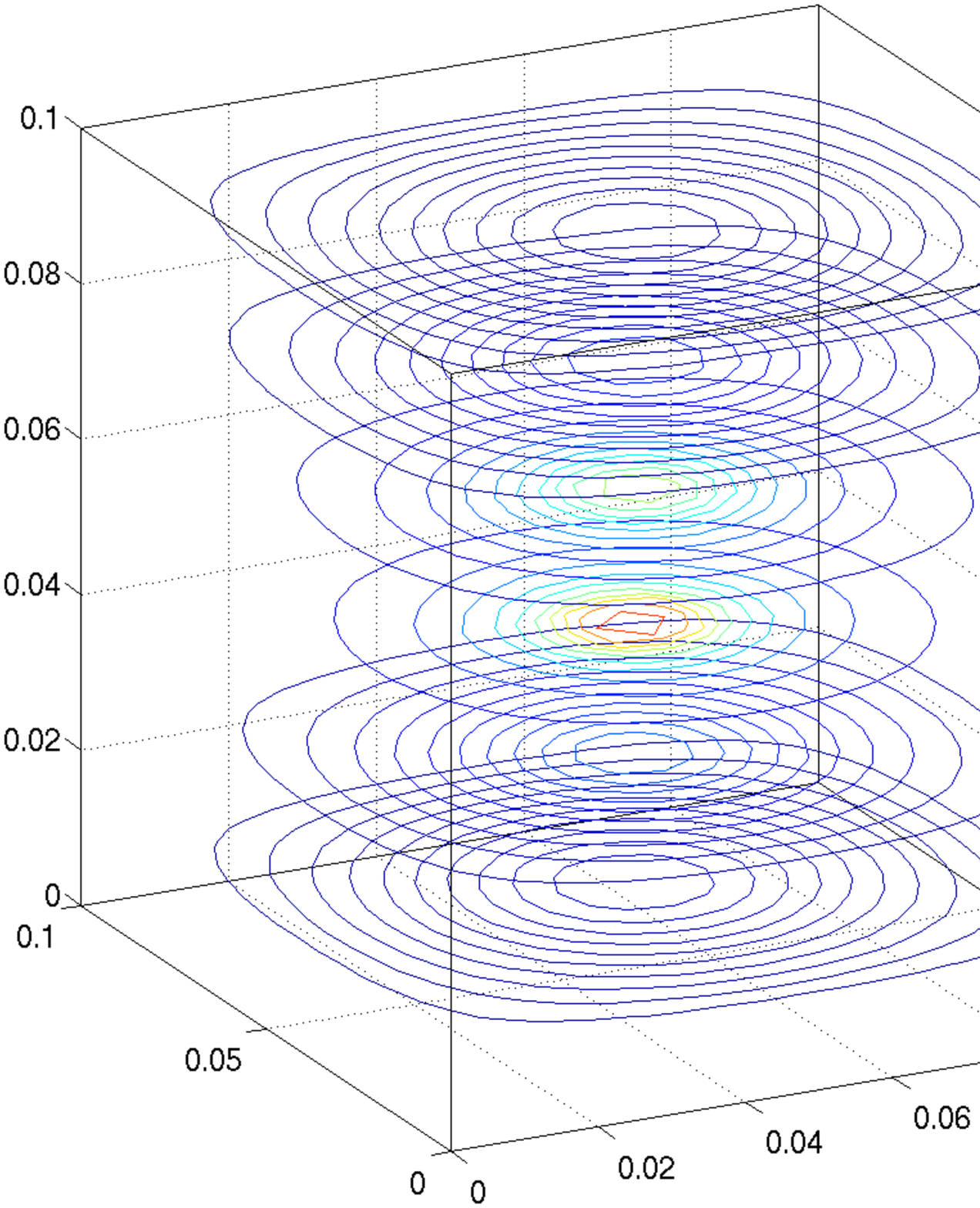}\label{fig:FEMcontourS1}}\\
	\caption{Slice  and contour comparison of solutions obtained using LFM and FEM for the source Fig\ref{fig:DBSexamples}(a). (a) Solution slices obtained with LFM. (b) Solution slices obtained with FEM. (c) Contours of solution obtained with LFM. (d) Contours of  solution obtained with FEM. }
	\label{fig:DiProb1}
\end{figure}

The mean of the conditional posterior distribution over the electric potential for the first source configuration (Fig. \ref{fig:DBSexamples}(a)), obtained through equation \eqref{e:f_posterior} using the proposed latent force model
approach, is showed in Fig. \ref{fig:LFMslicesS1} and \ref{fig:LFMcontourS1}. The corresponding electric potential, calculated using FEM for solving Poisson equation  is presented in Fig. \ref{fig:FEMslicesS1} and \ref{fig:FEMcontourS1}. There is a high similarity in shape as well as in magnitude compared with the LFM solution. 

\begin{figure}[!ht]
	\centering
	%\includegraphics[width=1\columnwidth]{model.eps}\\
	\subfigure[]{\includegraphics[width=0.33\columnwidth]{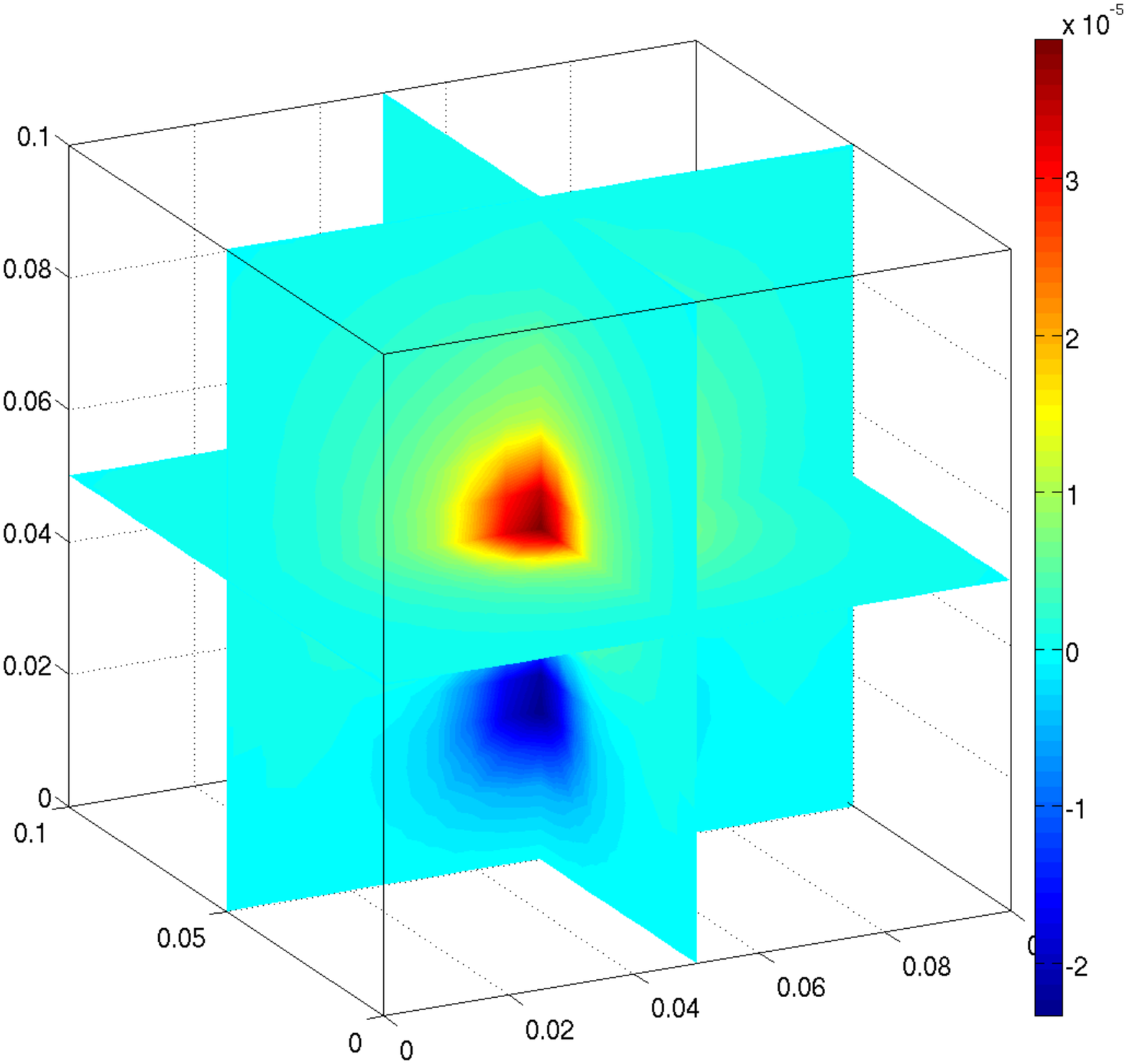}\label{fig:LFMslicesS2}}%\\
	\subfigure[]{\includegraphics[width=0.33\columnwidth]{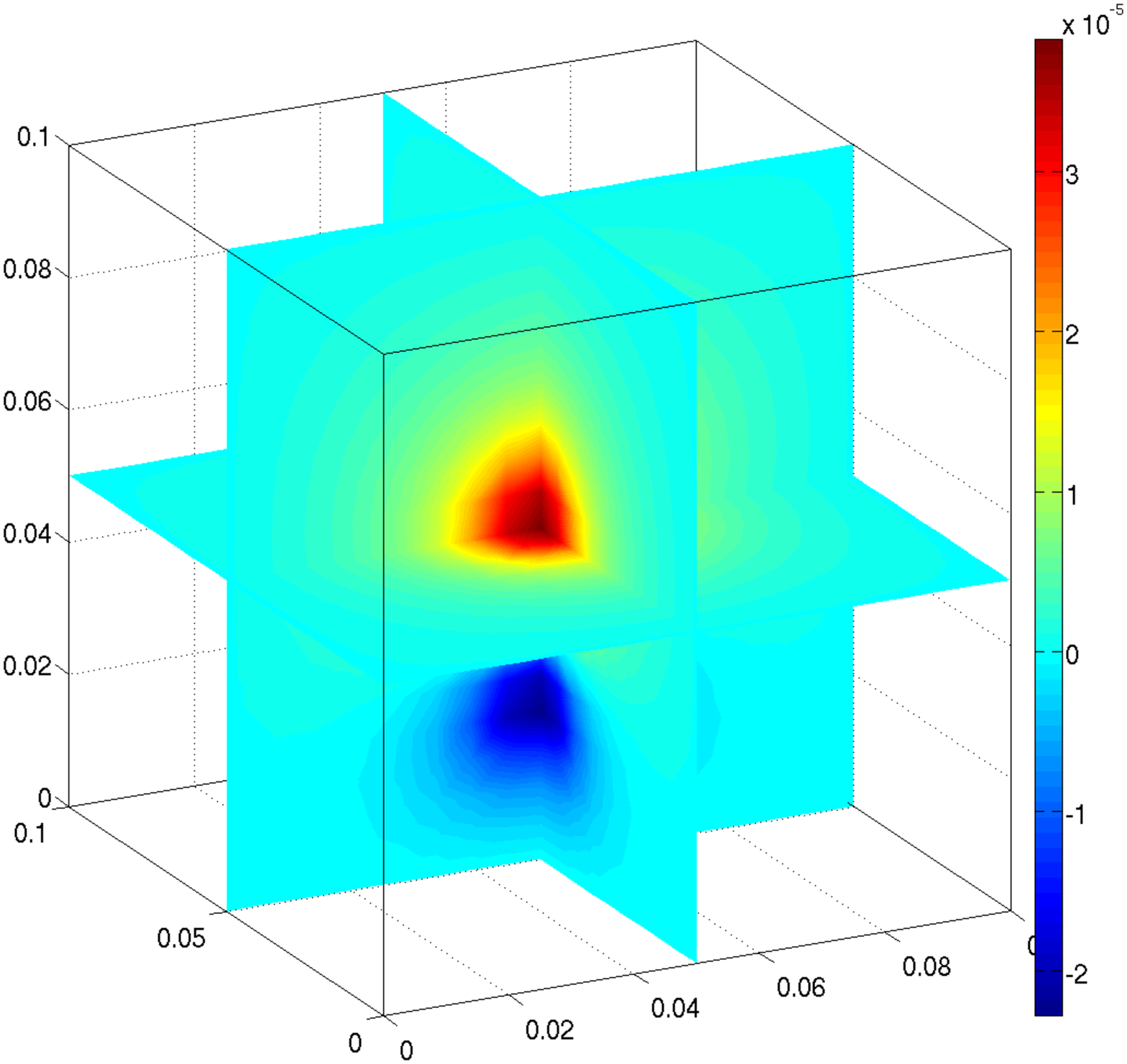}\label{fig:FEMslicesS2}}\\
	\subfigure[]{\includegraphics[width=0.33\columnwidth]{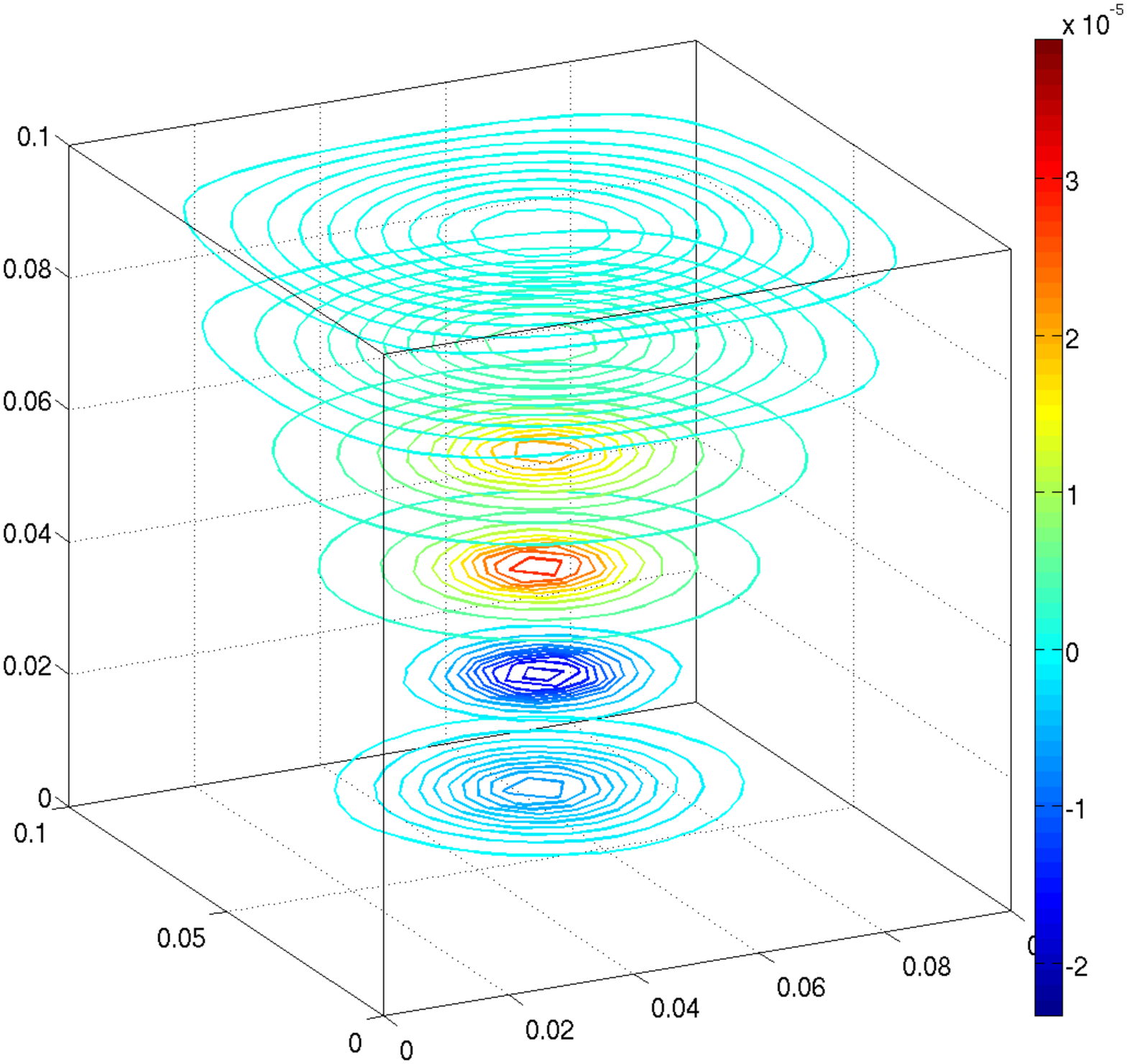}\label{fig:LFMcontourS2}}%\\
	\subfigure[]{\includegraphics[width=0.33\columnwidth]{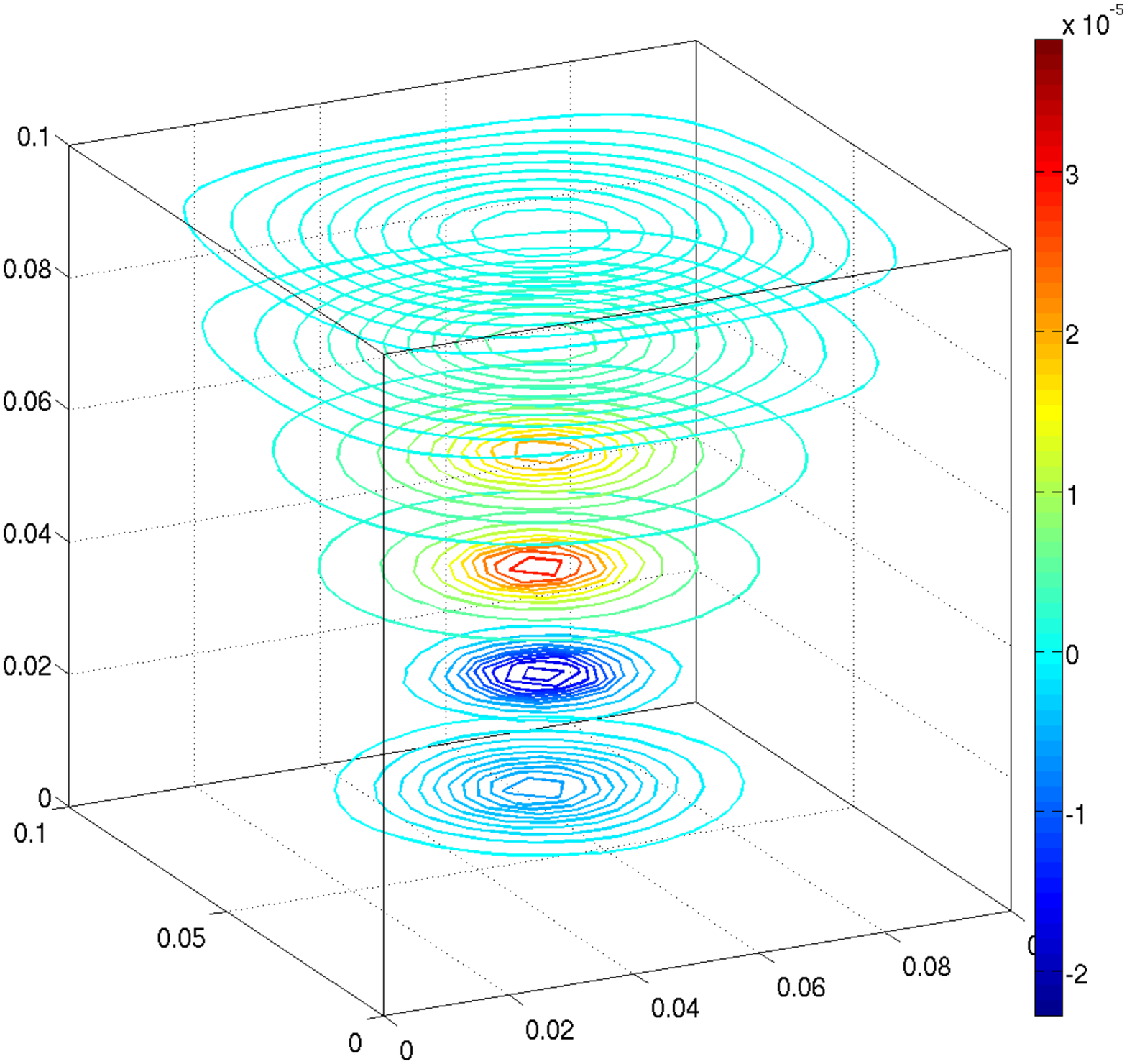}\label{fig:FEMcontourS2}}\\
	\caption{Slice  and contour comparison of solutions obtained using LFM and FEM for the source Fig\ref{fig:DBSexamples}(b). (a) Solution slices obtained with LFM. (b) Solution slices obtained with FEM. (c) Contours of solution obtained with LFM. (d) Contours of solution obtained with FEM. }
	\label{fig:DiProb2}
\end{figure}

\begin{figure}[h!]
	\centering
	%\includegraphics[width=1\columnwidth]{model.eps}\\
	\subfigure[]{\includegraphics[width=0.33\columnwidth]{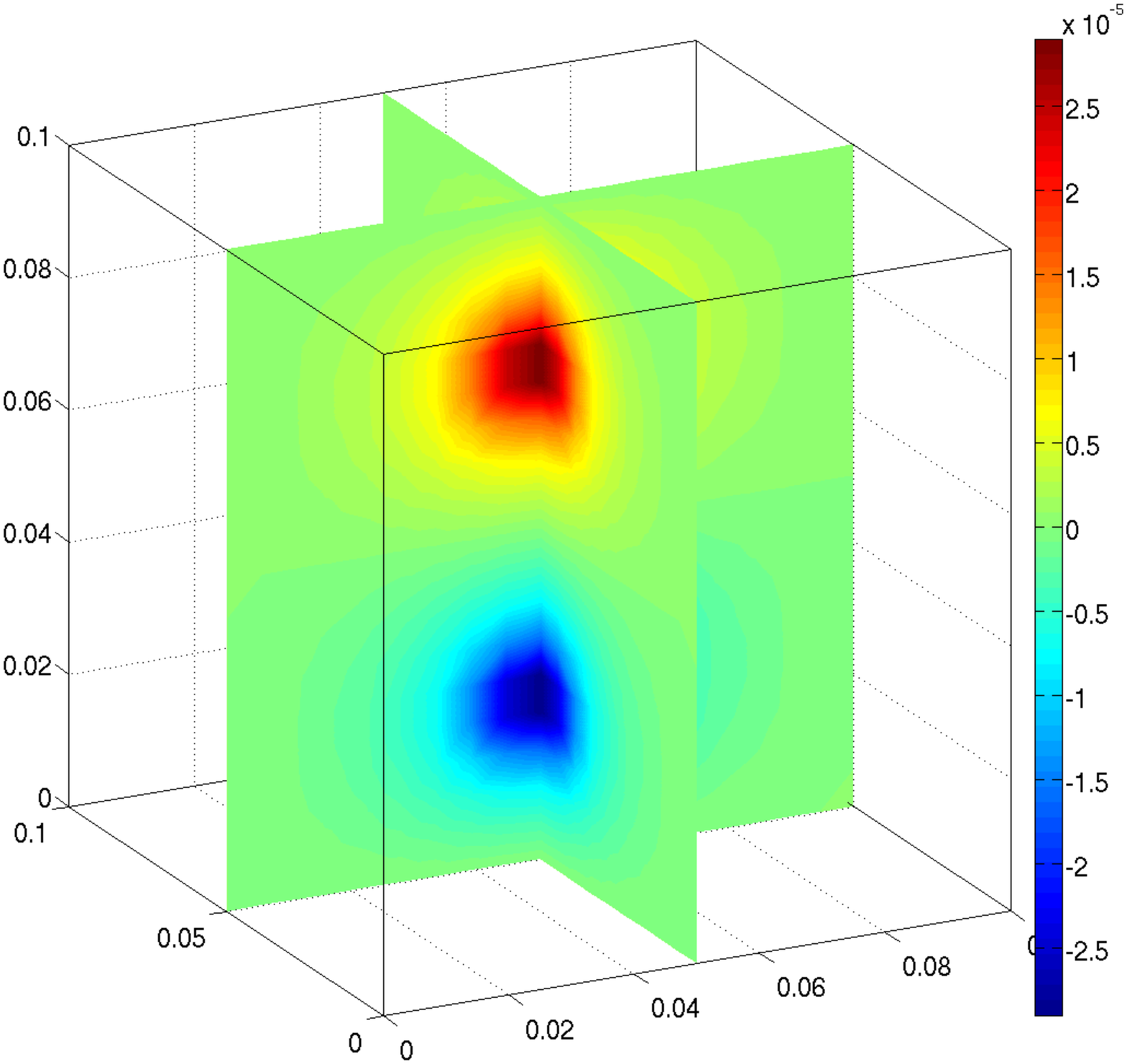}\label{fig:LFMslicesS3}}%\\
	\subfigure[]{\includegraphics[width=0.33\columnwidth]{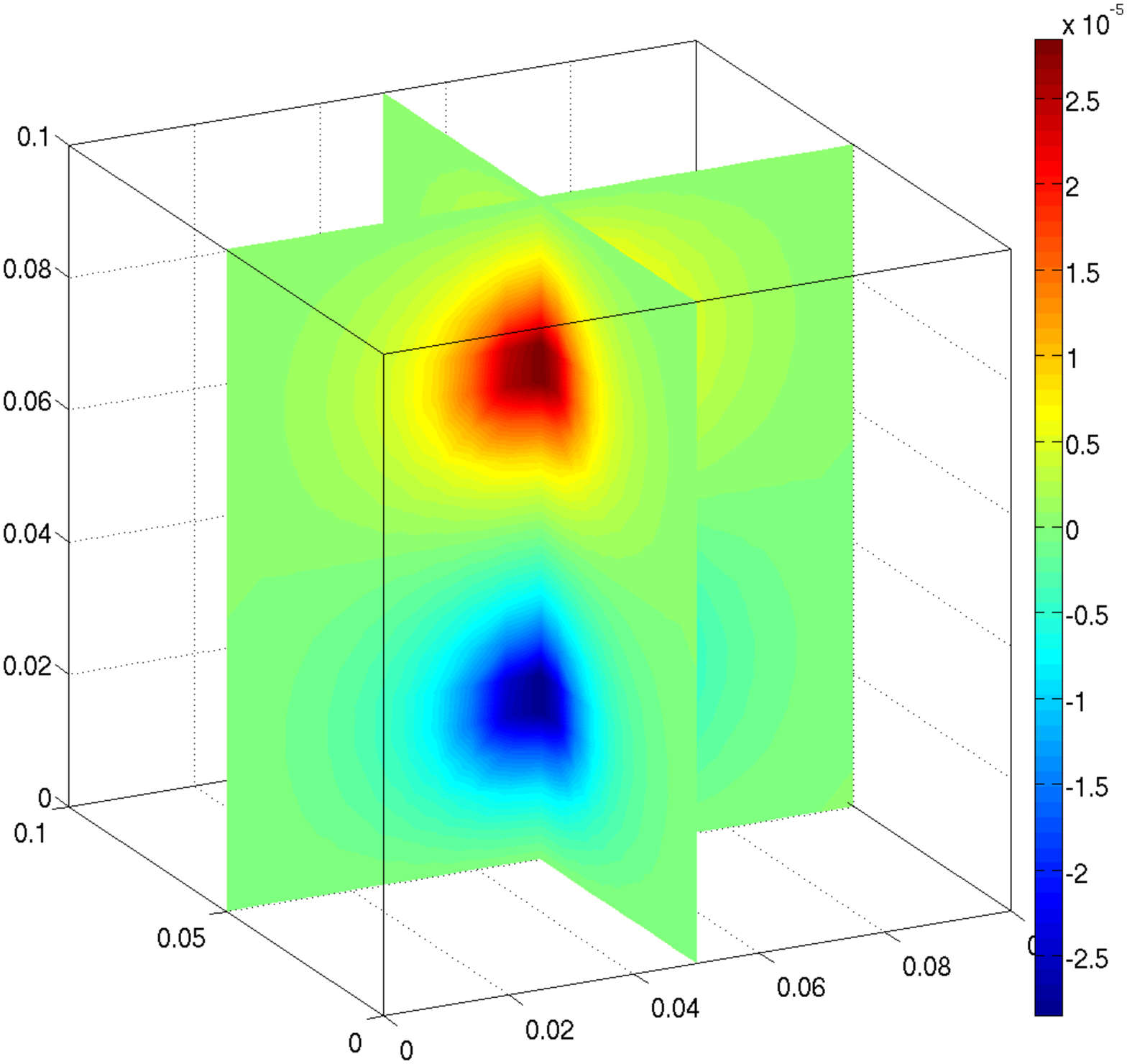}\label{fig:FEMslicesS3}}\\
	\subfigure[]{\includegraphics[width=0.33\columnwidth]{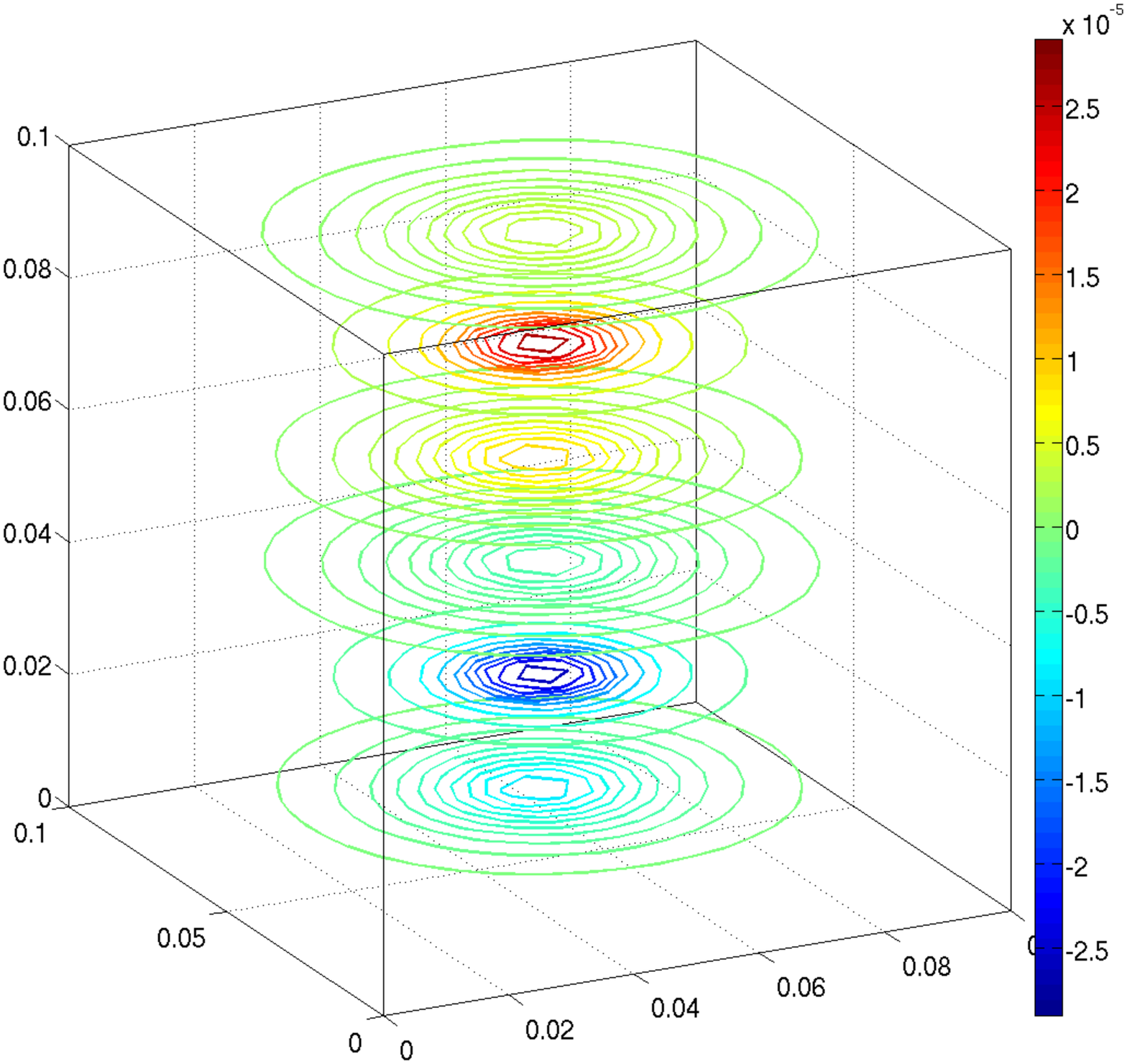}\label{fig:LFMcontourS3}}%\\
	\subfigure[]{\includegraphics[width=0.33\columnwidth]{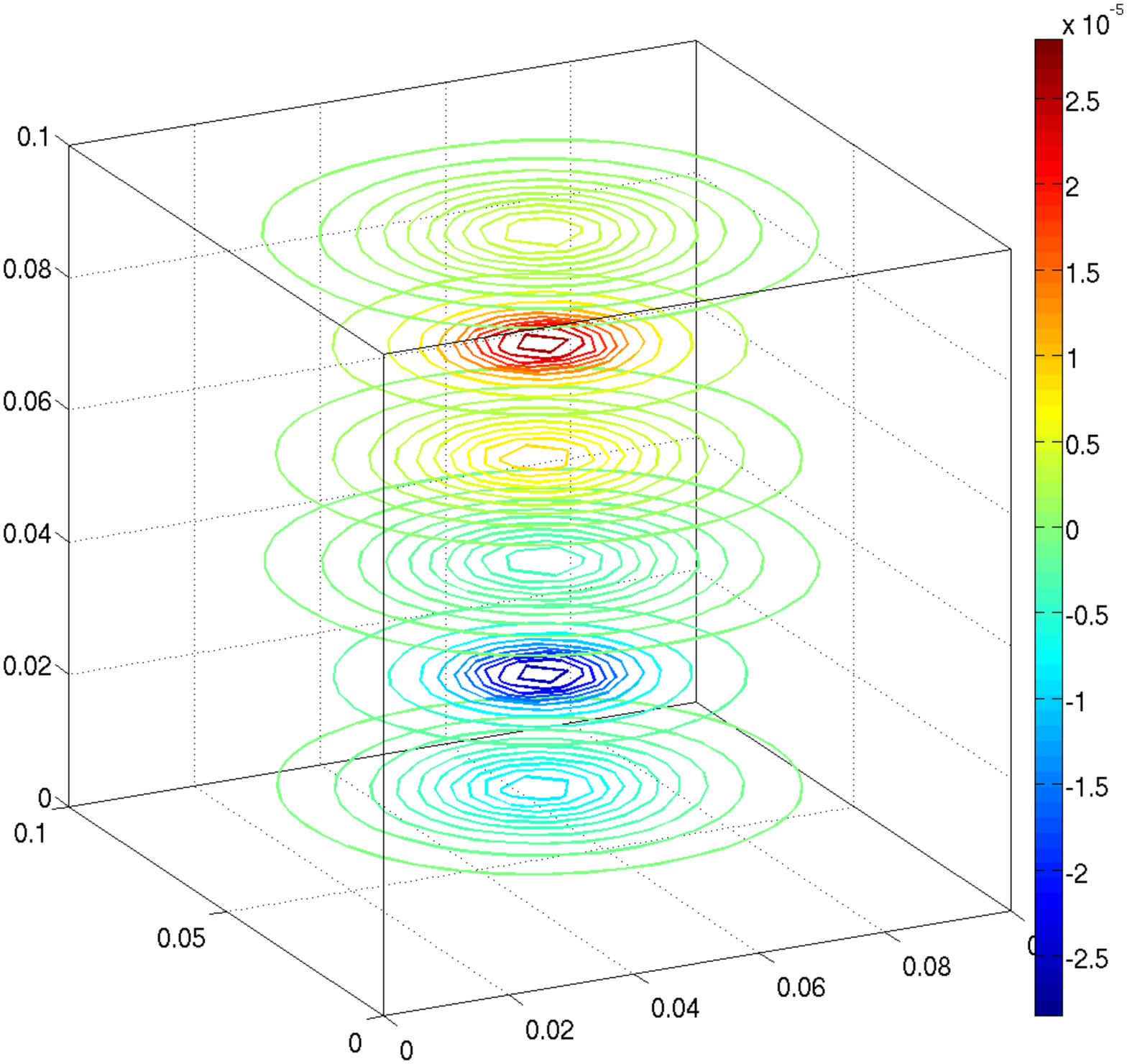}\label{fig:FEMcontourS3}}\\
	\caption{Slice  and contour comparison of solutions obtained using LFM and FEM for the source Fig\ref{fig:DBSexamples}(c). (a) Solution slices obtained with LFM. (b) Solution slices obtained with FEM. (c) Contours of solution obtained with LFM. (d) Contours of solution obtained with FEM. }
	\label{fig:DiProb3}
\end{figure}

The electric potential for the second source configuration (Fig. \ref{fig:DBSexamples}(b)), calculated using the proposed latent force model
approach is showed in Fig. \ref{fig:LFMslicesS2} and \ref{fig:LFMcontourS2}. The corresponding electric potential, calculated using FEM for solving Poisson equation  is presented in Fig. \ref{fig:FEMslicesS2} and \ref{fig:FEMcontourS2}.
The posterior mean over the electric potential for the third source configuration (Fig. \ref{fig:DBSexamples}(c)), obtained with the wave latent force model is showed in Fig. \ref{fig:LFMslicesS3} and \ref{fig:LFMcontourS3}. The corresponding electric potential, calculated using FEM for solving the Poisson equation  is presented in Fig. \ref{fig:FEMslicesS3} and \ref{fig:FEMcontourS3}. For both cases, similar to the first electrode configuration results (Fig.\ref{fig:DiProb1}), the LFM solutions are close to the corresponding solution obtained using FEM.

\begin{figure}[h!]
	\centering
	\includegraphics[width=0.33\columnwidth]{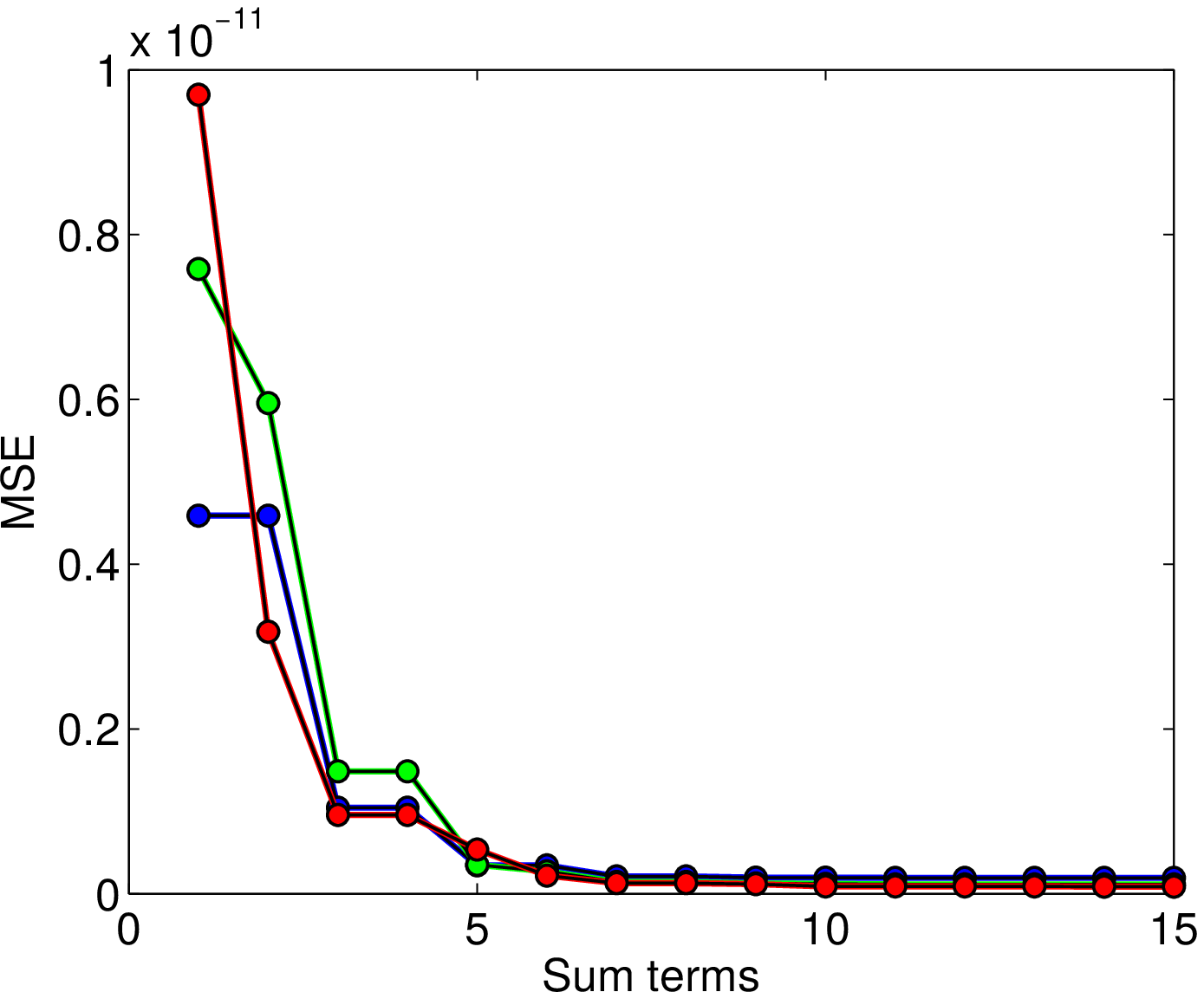}
	\caption{Mean squared error (MSE) between the solution obtained with FEM and LFM, for different number of terms in the Green's function \eqref{e:Green}. %This figure holds for a different value for each hyperparameter, they were tuned manually. 
	(Blue) results for source Fig. \ref{fig:DBSexamples}(a), (green) results for source Fig. \ref{fig:DBSexamples}(b), (red) results for source Fig. \ref{fig:DBSexamples}(c).}
	\label{fig:meanerror}
\end{figure}

\begin{figure}[!ht]
	\centering
	\subfigure[]{\includegraphics[width=0.33\columnwidth]{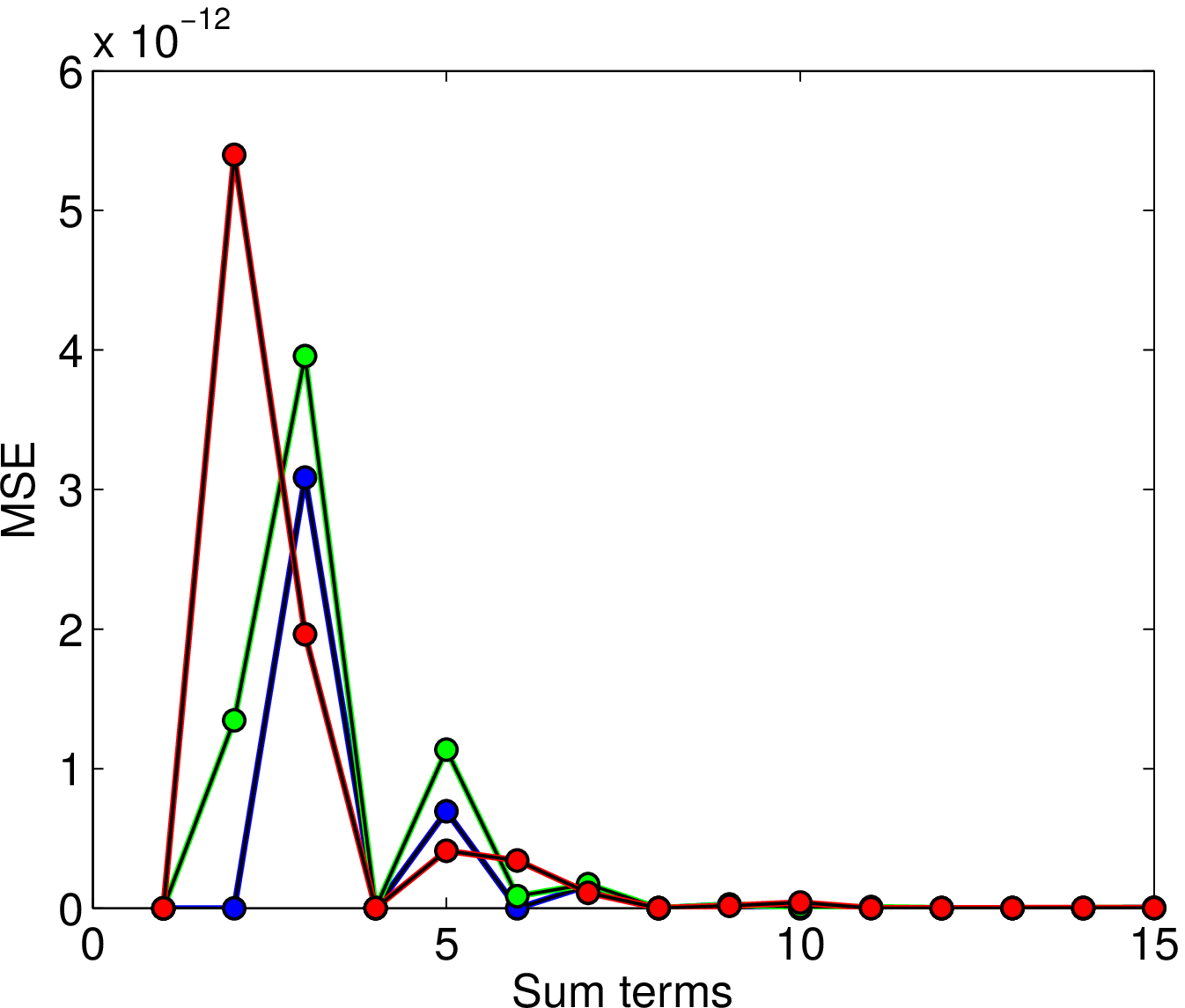}\label{fig:meandif}}
	\subfigure[]{\includegraphics[width=0.33\columnwidth]{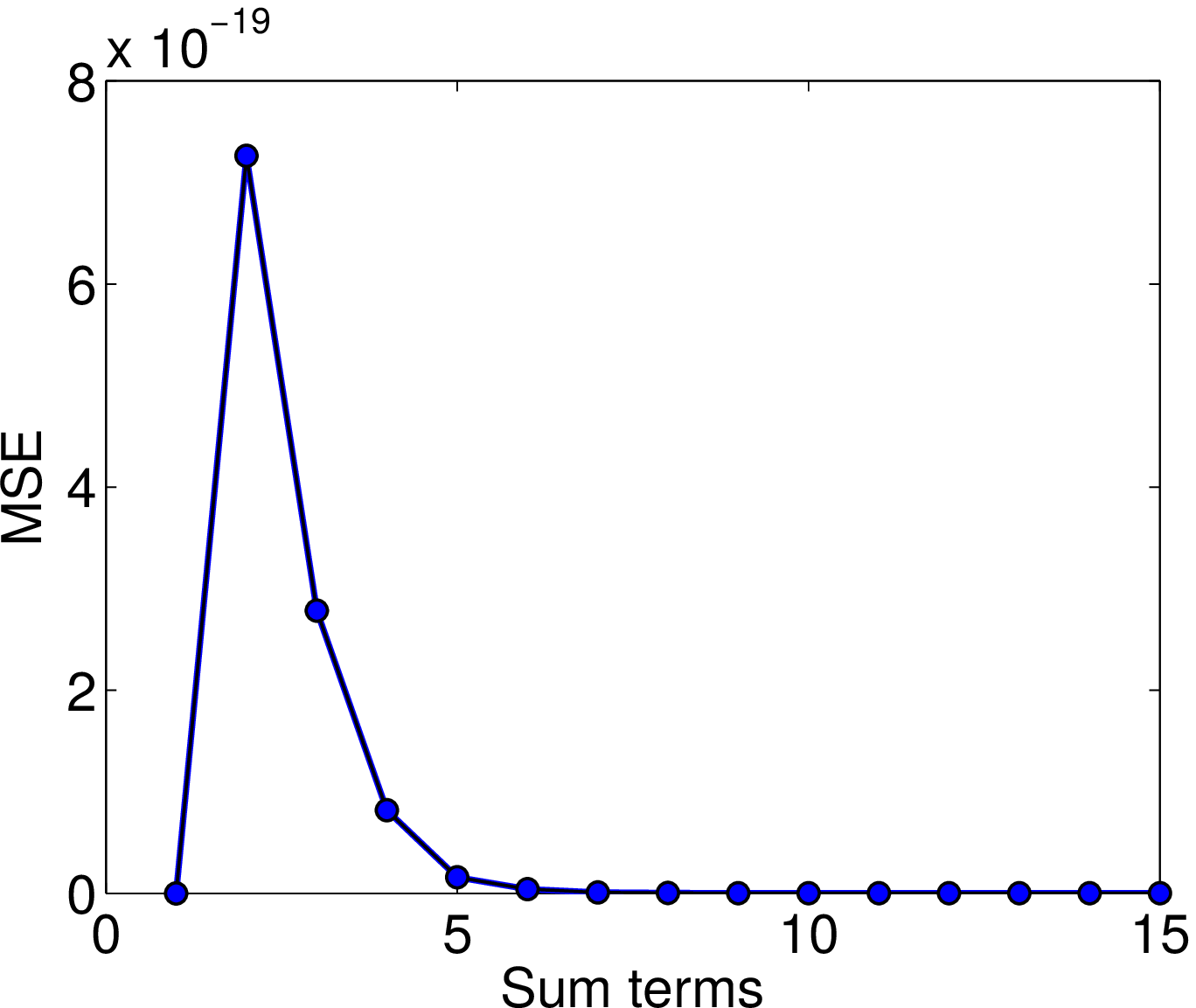}\label{fig:changePriorvar}}\\
	\subfigure[]{\includegraphics[width=0.33\columnwidth]{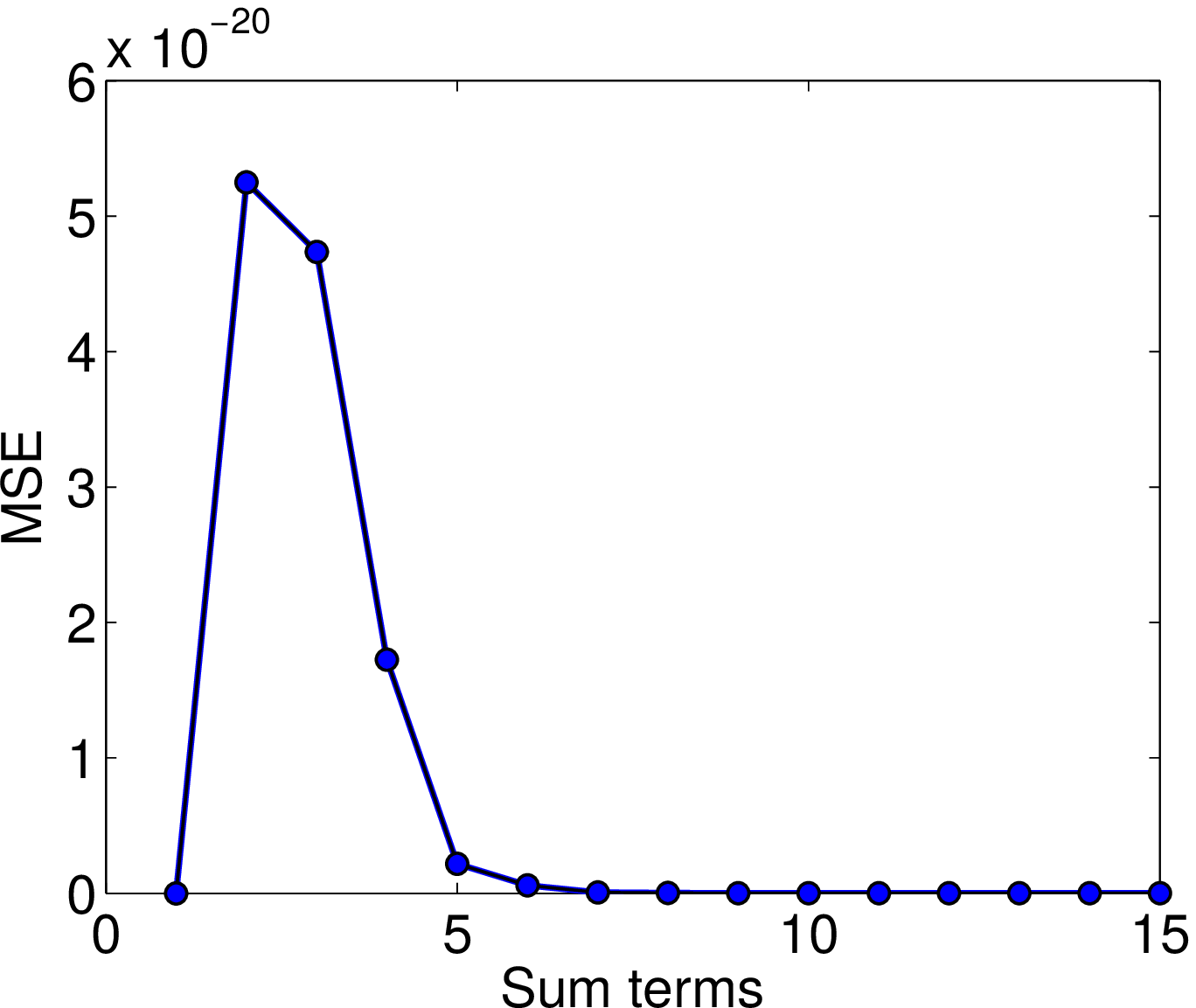}\label{fig:changePostvar}}\\
	\caption{ Variation in the posterior mean, prior variance, and posterior variance, calculated for different number of terms in the sums of the Green's function \eqref{e:Green}. Each point represents the variation  between the results of using one term and the results of using two terms, then the variation between using two terms and three, and so on.  (a) Variation between consecutive posterior mean, (blue) results for source Fig. \ref{fig:DBSexamples}(a), (green) results for source Fig. \ref{fig:DBSexamples}(b), (red) results for source Fig. \ref{fig:DBSexamples}(c) . (b) Variation between consecutive prior variances. (c) Variation between consecutive posterior variances.}
	\label{fig:allDif}
\end{figure}

The covariance function  $k_f(\textbf{x},\textbf{x}';t,t')$ of the output,  as well as  the cross covariance function  $k_{fu}(\textbf{x},\textbf{x}';t,t')$ between the latent function and the solution of the wave equation, depend both on the number of terms used for each sum in the expression for the Green's function \eqref{e:Green}. %Figures \ref{fig:meantime} and \ref{fig:timepostvariancer} show the time used for calculating the posterior mean and posterior variance of the distribution over the solution of the wave equation. It can be seen that the computational cost increases with the number of terms used in \eqref{e:Green}.
Figure \ref{fig:meanerror} presents the mean squared error between the solution obtained with FEM and LFM for the three sources simulated in this section, for different numbers of terms in the sums needed for the computation of the posterior mean over the solution function of the wave equation. This Figure suggests that with approximately seven terms in each of the three sums in \eqref{e:Green} we can obtain an appropriate approximation.
%
% cambio
%\begin{figure}[h!]
%	\centering
%	\subfigure[]{\includegraphics[width=0.495\columnwidth]{imagenes/mean_time}\label{fig:meantime}}%\\
%	\subfigure[]{\includegraphics[width=0.495\columnwidth]{imagenes/timepostvariance}\label{fig:timepostvariancer}}%\\
%	\caption{(a) Time (in seconds) needed for computing the posterior mean. (b) Time (in seconds) needed for computing the posterior variance. }
%	\label{fig:times}
%\end{figure} 
%
%
Finally,
to show how the number of terms in the sums present in the Green's function \eqref{e:Green} affects the results, we calculate the variation between results using different number of terms in the sums. Each point in Figure \ref{fig:allDif} represents the variation  between the results of using one term and the results of using two terms, then the variation between using two terms and three, and so on. 
%
%entre diferentes resultados utilizando un determinada cantidad de terminos en la sumas. las figuras tal, tal y tal muestran la variacion entre usar un termino en las sumas y usar dos terminos, despues la diferencia entre usar dos terminos y tres terminos, y asi consecutivamente....
%
Fig. \ref{fig:meandif}, \ref{fig:changePriorvar} and \ref{fig:changePostvar} show the variation in the posterior mean over the electric potential obtained for each source in Fig. \ref{fig:DBSexamples}, as well as the prior and posterior variance, calculated for different number of terms in the solution sum. This information also allows us to conclude that with approximately seven terms in the sums in \eqref{e:Green}  we can obtain a good approximation. 

\subsection{Inverse Problem Approach}
 \noindent
 The wave LFM was used for recovering four different DBS excitation configurations. The corresponding electric potentials calculated using FEM were the input data of the wave LFM in equation \eqref{e:u_posterior} to get the conditional posterior
 distributions over the sources. Every electrode configuration was approximated with a mixture of Gaussian distributions. For illustrating these results only two spatial dimensions were used, i.e. $\textbf{x}\in \mathbb{R}^{2}$.
 %
% \begin{figure}[!Ht]  
% 	\centering
% 	\subfigure {\def\svgwidth{150pt} \input{imagenes/DBSlead/leadInverseProblem.eps_tex}}
% 	\caption{Electrode configurations used for solving the inverse problem.}
% 	\label{fig:DBSexamplesInvProb}
% \end{figure}
 %
 
\begin{figure}[!ht]
 	\centering
 	\begin{tabular}{cc}
 	\subfigure[]{\includegraphics[width=0.02\columnwidth]{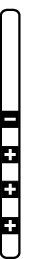}\label{fig:InvProbLead1}} 
 	&
 	\subfigure[]{\includegraphics[width=0.33\columnwidth]{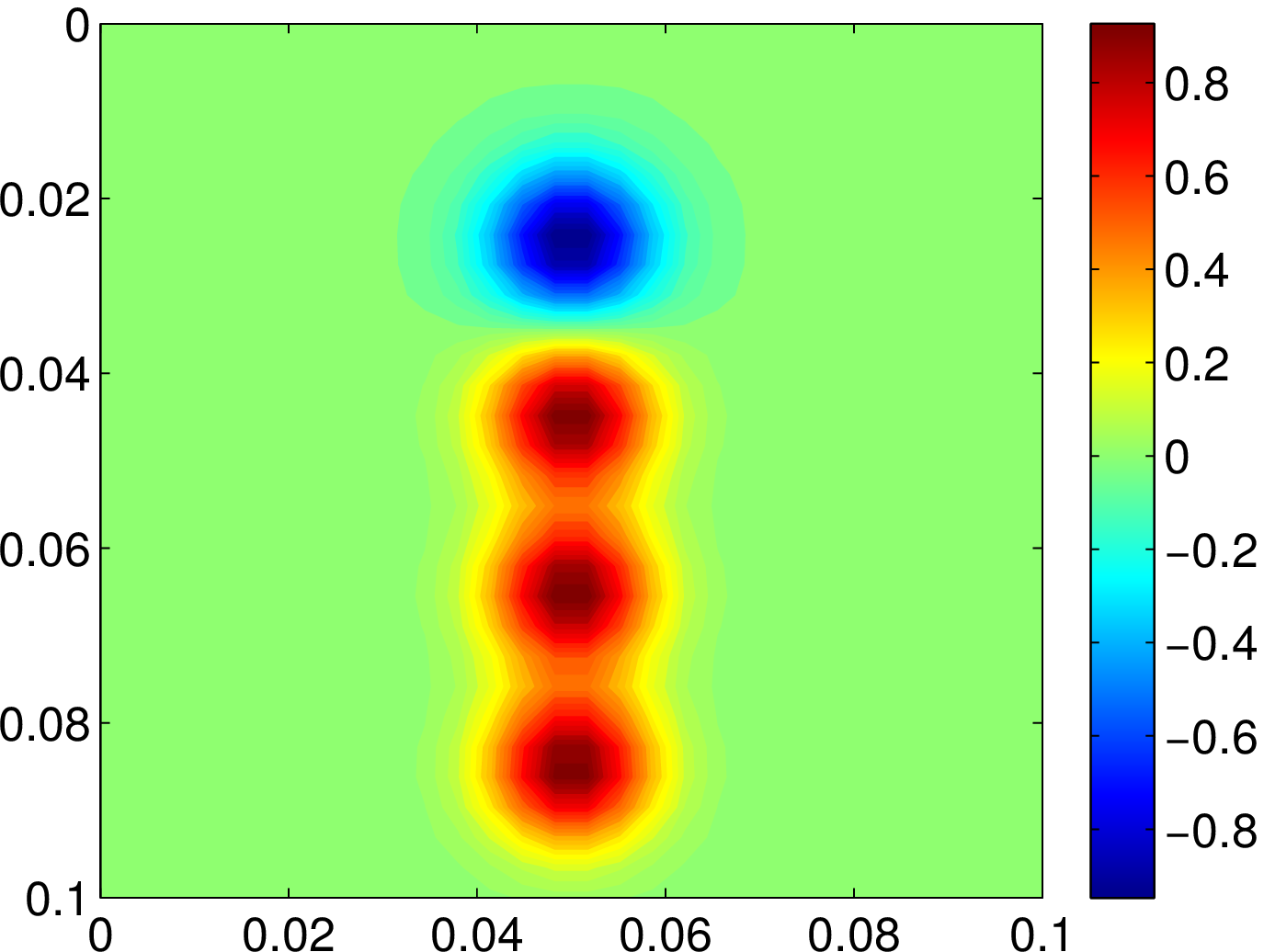}\label{fig:ip_source1}} 
 	\\
 	\subfigure[]{\includegraphics[width=0.33\columnwidth]{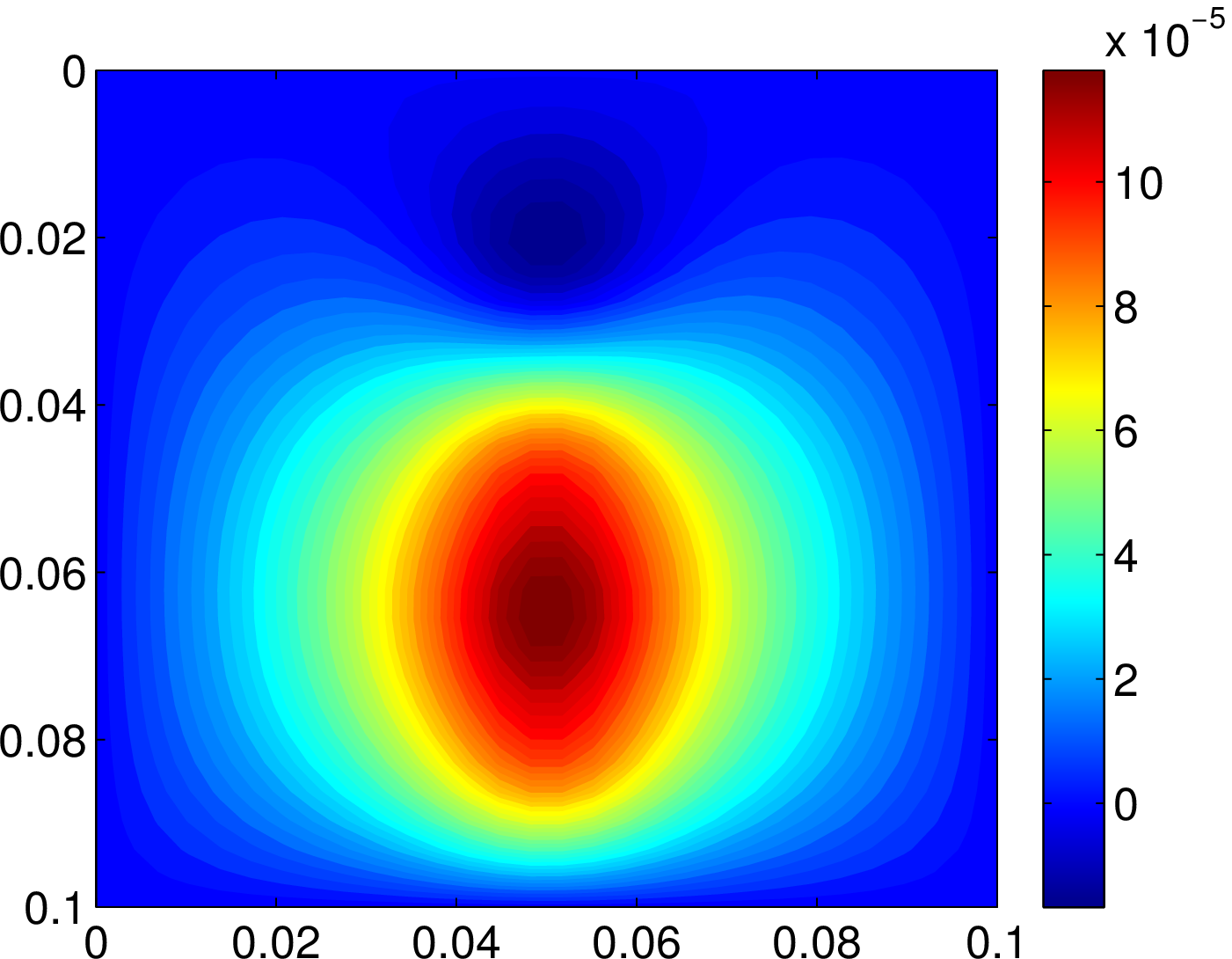}\label{fig:ip_potential1}} 
 	&
 	\subfigure[]{\includegraphics[width=0.33\columnwidth]{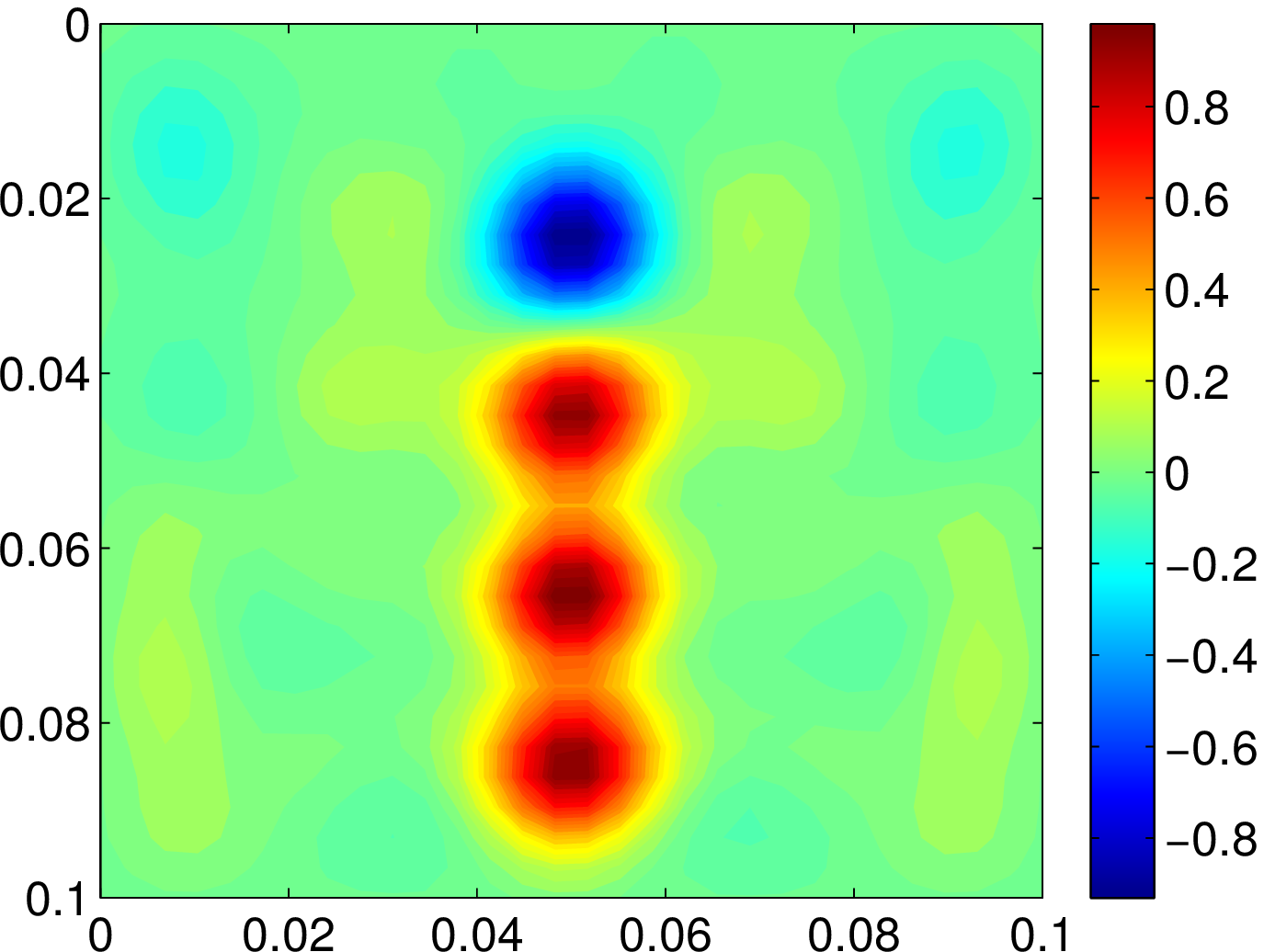}\label{fig:ip_LFMsource1}} 
    \end{tabular}
 	\caption{ (a) electrode configuration, (b) its corresponding source of excitation function. (c) electric potential produced. (d) the source recovered using the LFM, with MSE  $= 2.5\times 10^{-3}$.}
 	\label{fig:InvProbResults1}
 \end{figure}

\begin{figure}[!ht]
	 	\centering
	 	\begin{tabular}{cc}
 	\subfigure[]{\includegraphics[width=0.02\columnwidth]{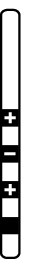}\label{fig:InvProbLead2}} 
 	&
 	\subfigure[]{\includegraphics[width=0.33\columnwidth]{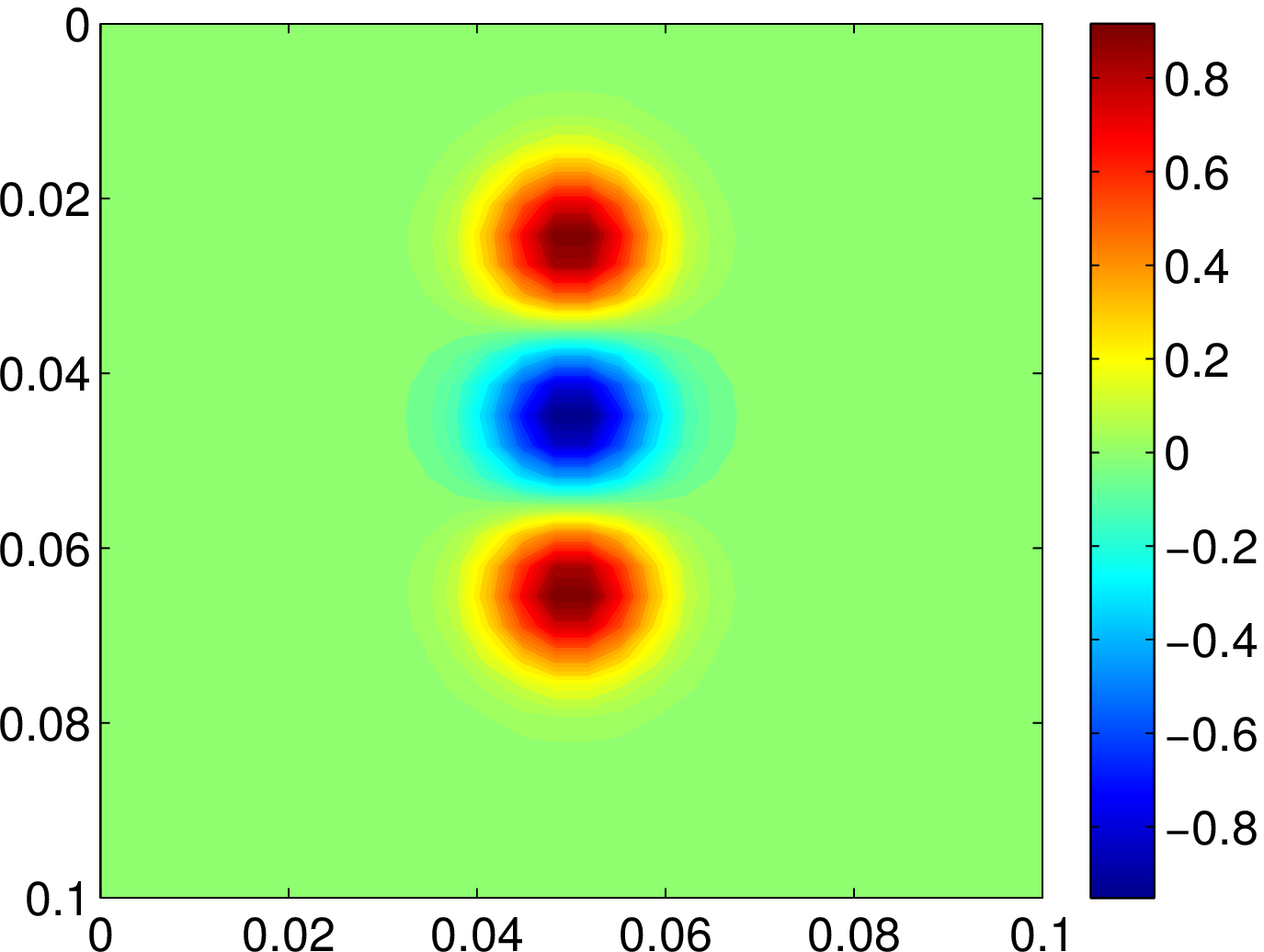}\label{fig:ip_source2}} 
 	\\
 	\subfigure[]{\includegraphics[width=0.33\columnwidth]{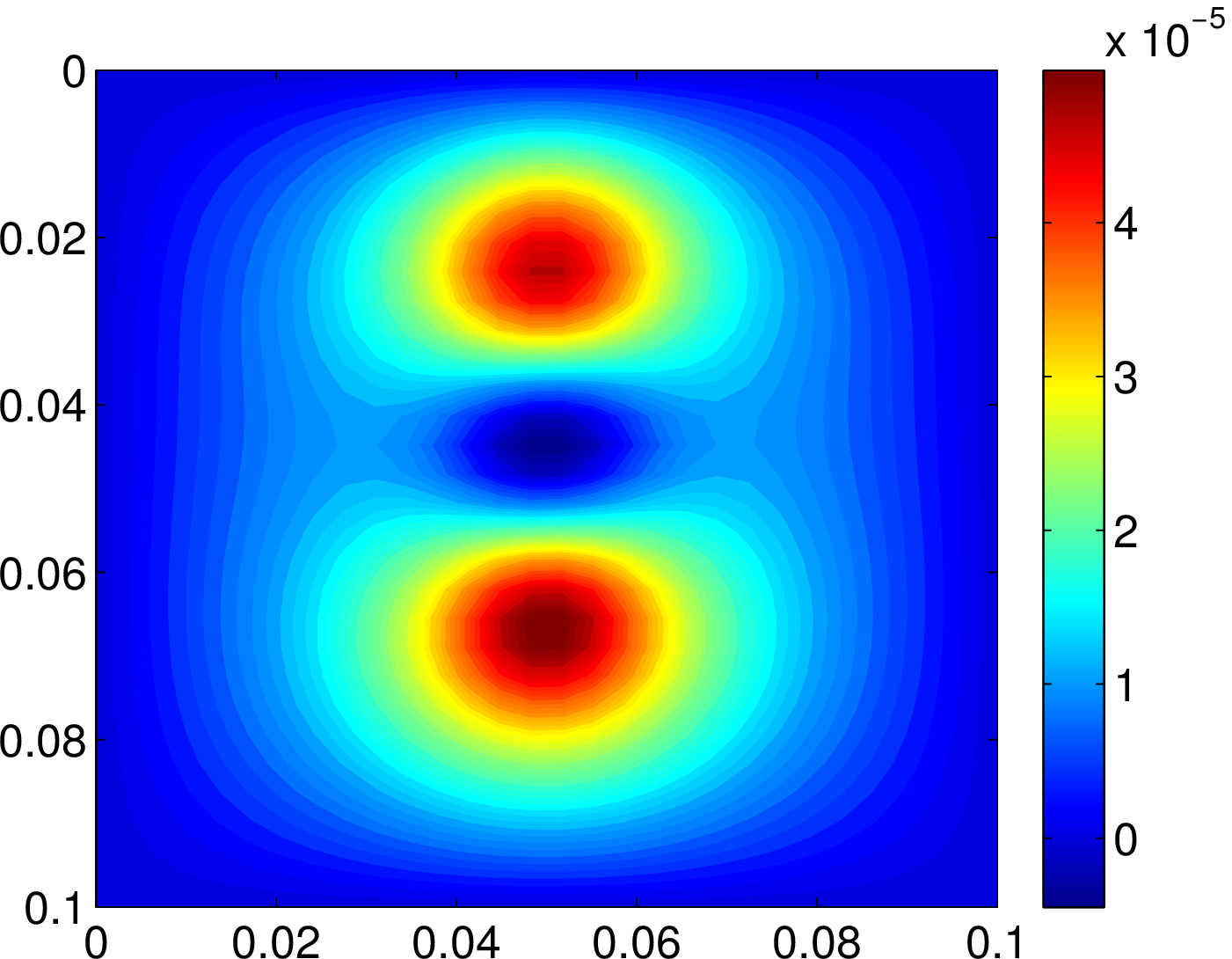}\label{fig:ip_potential2}} 
 	&
 	\subfigure[]{\includegraphics[width=0.33\columnwidth]{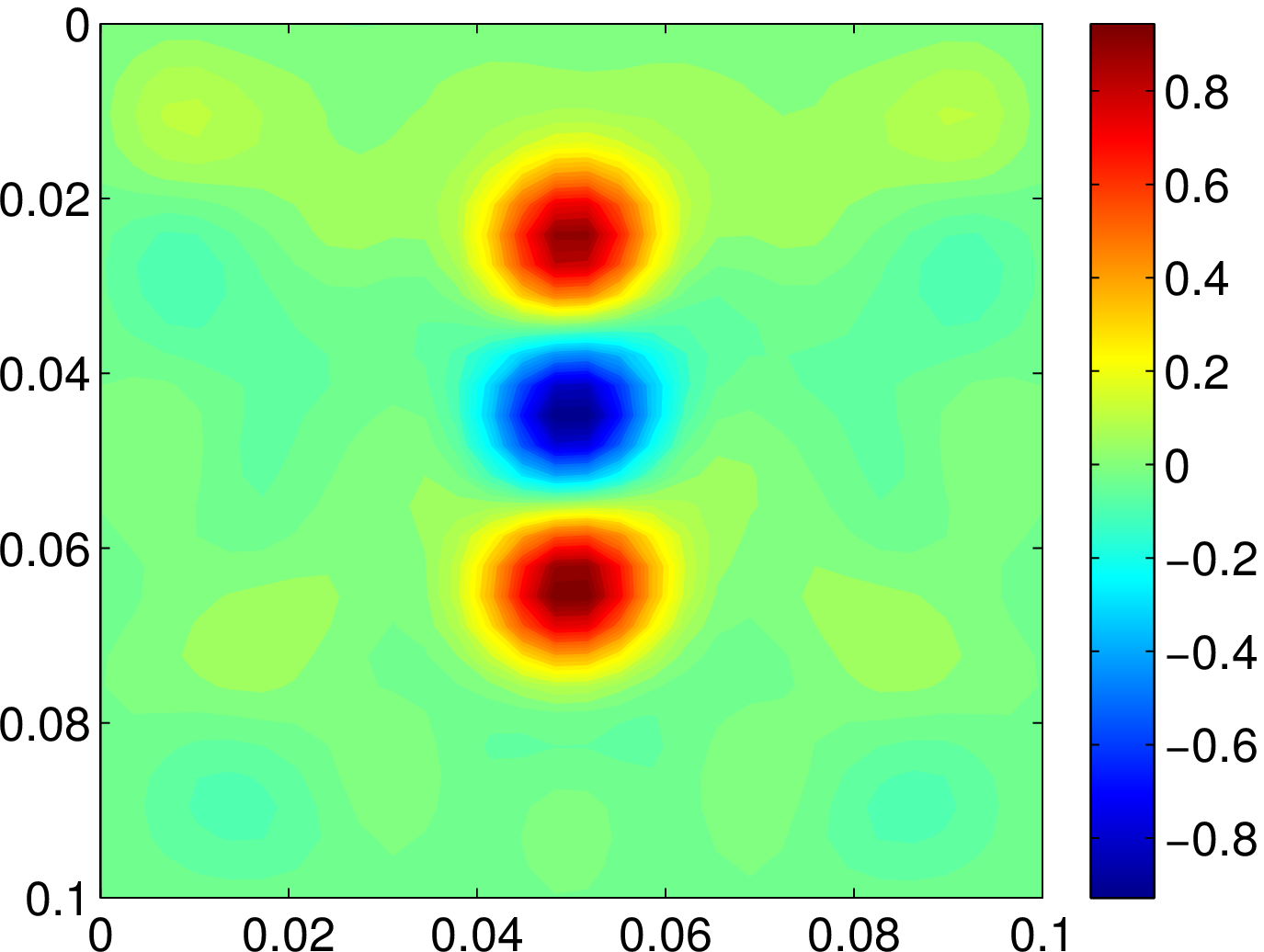}\label{fig:ip_LFMsource2}}  	
	\end{tabular}
 	\caption{ (a) electrode configuration, (b) its corresponding source of excitation function. (c) electric potential produced. (d) the source recovered using the LFM, with MSE $= 1.8\times 10^{-3}$.}
 	\label{fig:InvProbResults2}
 \end{figure}

\begin{figure}[!ht]
 	\centering
 	\begin{tabular}{cc}
 	\subfigure[]{\includegraphics[width=0.02\columnwidth]{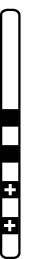}\label{fig:InvProbLead3}} 
 	&
 	\subfigure[]{\includegraphics[width=0.33\columnwidth]{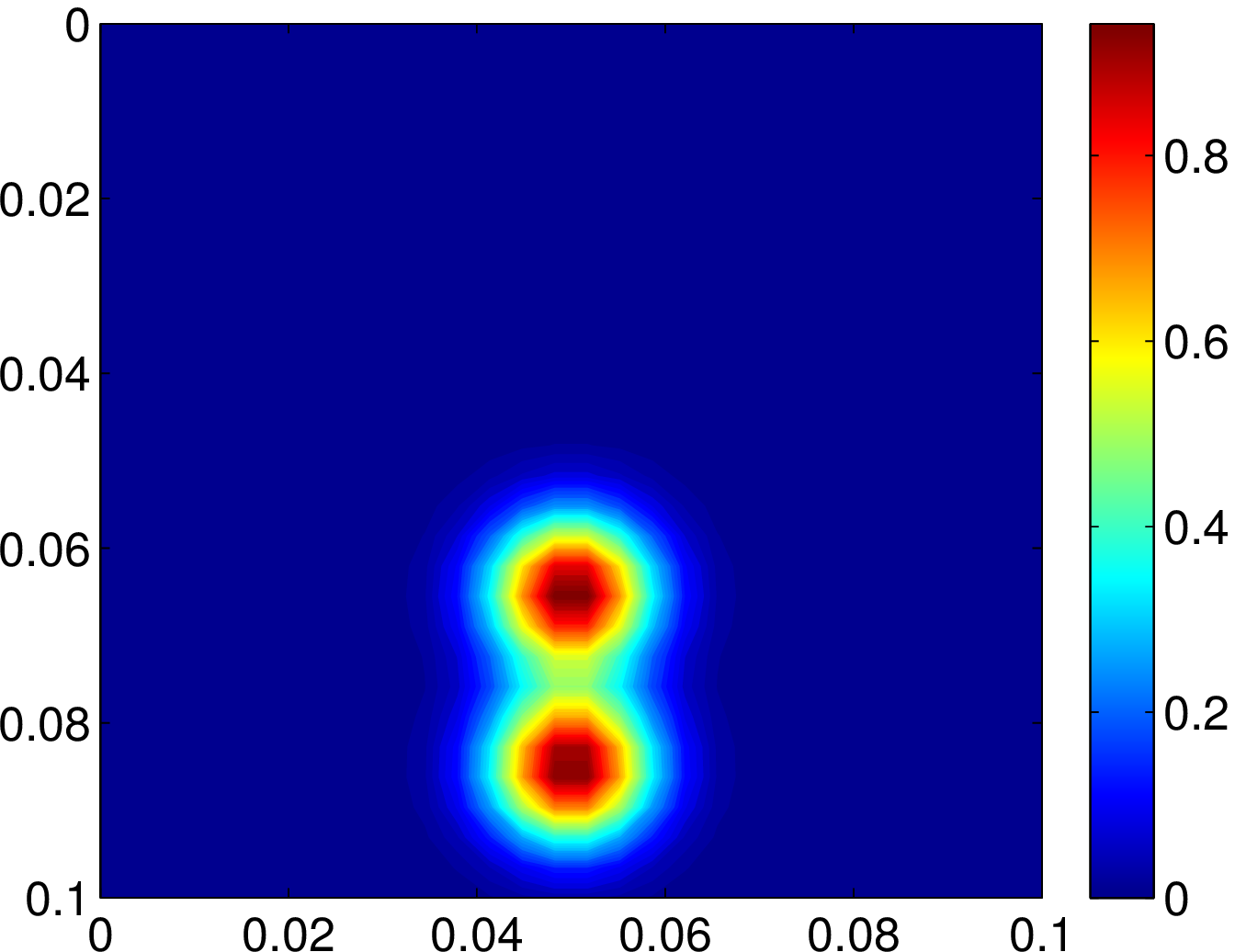}\label{fig:ip_source3}} 
 	\\
 	\subfigure[]{\includegraphics[width=0.33\columnwidth]{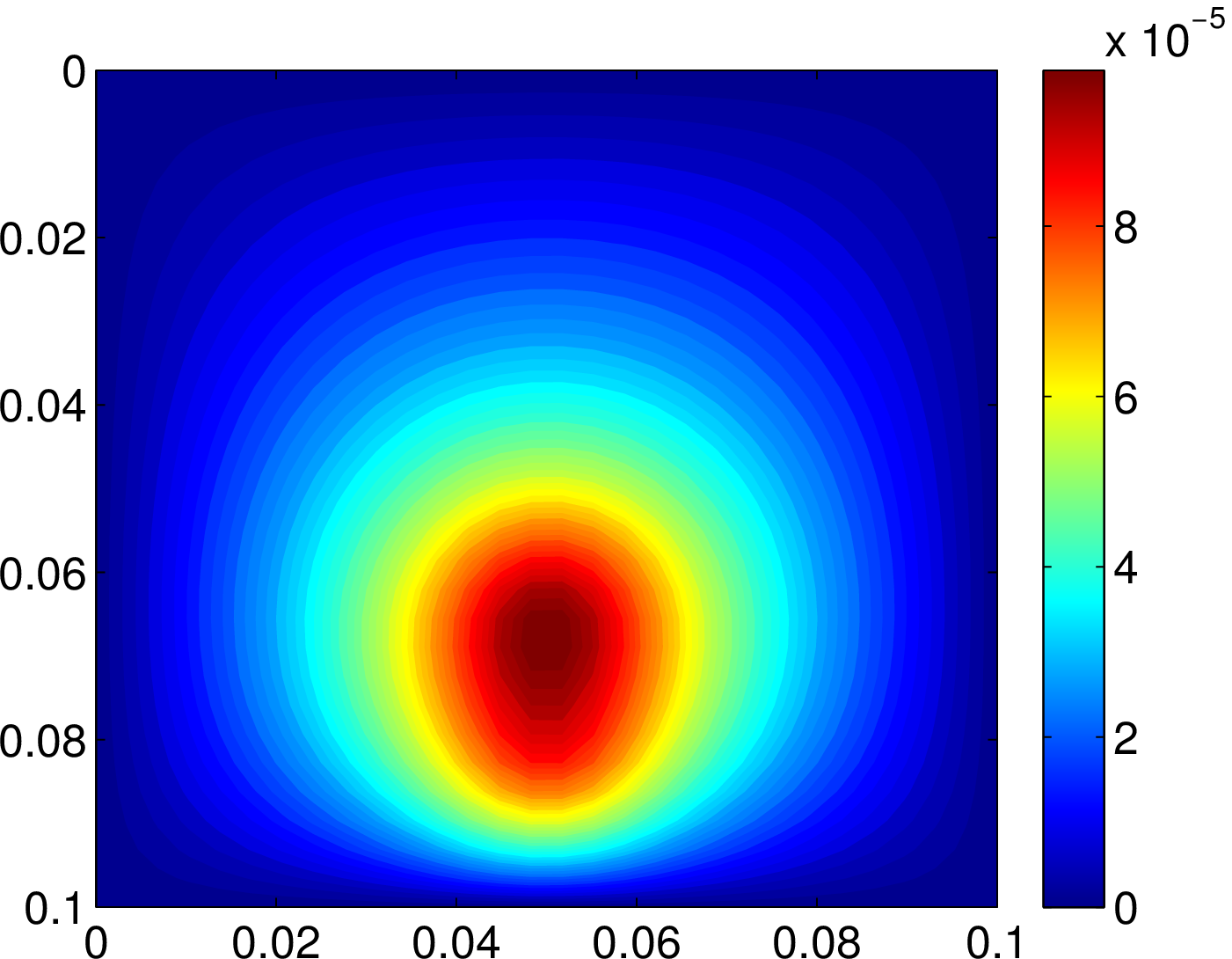}\label{fig:ip_potential3}} 
 	&
 	\subfigure[]{\includegraphics[width=0.33\columnwidth]{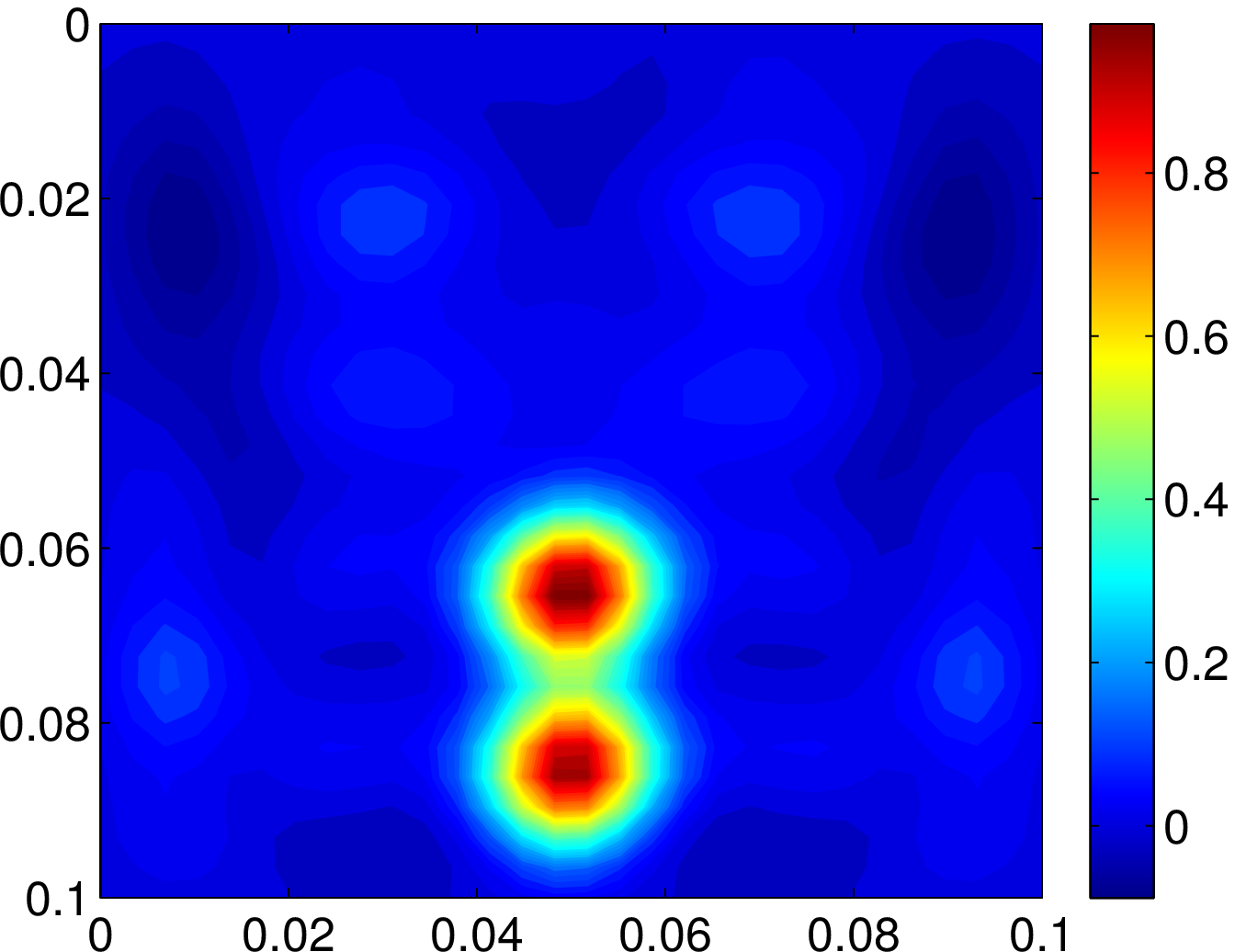}\label{fig:ip_LFMsource3}}
 	\end{tabular}  
 	
 	 	\caption{ (a) electrode configuration, (b) its corresponding source of excitation function. (c) electric potential produced. (d) the source recovered using the LFM, with MSE  $= 1.2\times 10^{-3}$.}
 	\label{fig:InvProbResults3}
 	\end{figure}

\begin{figure}
 			\centering
	\begin{tabular}{cc}
 	\subfigure[]{\includegraphics[width=0.02\columnwidth]{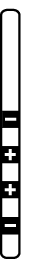}\label{fig:InvProbLead4}} 
 	&
 	\subfigure[]{\includegraphics[width=0.33\columnwidth]{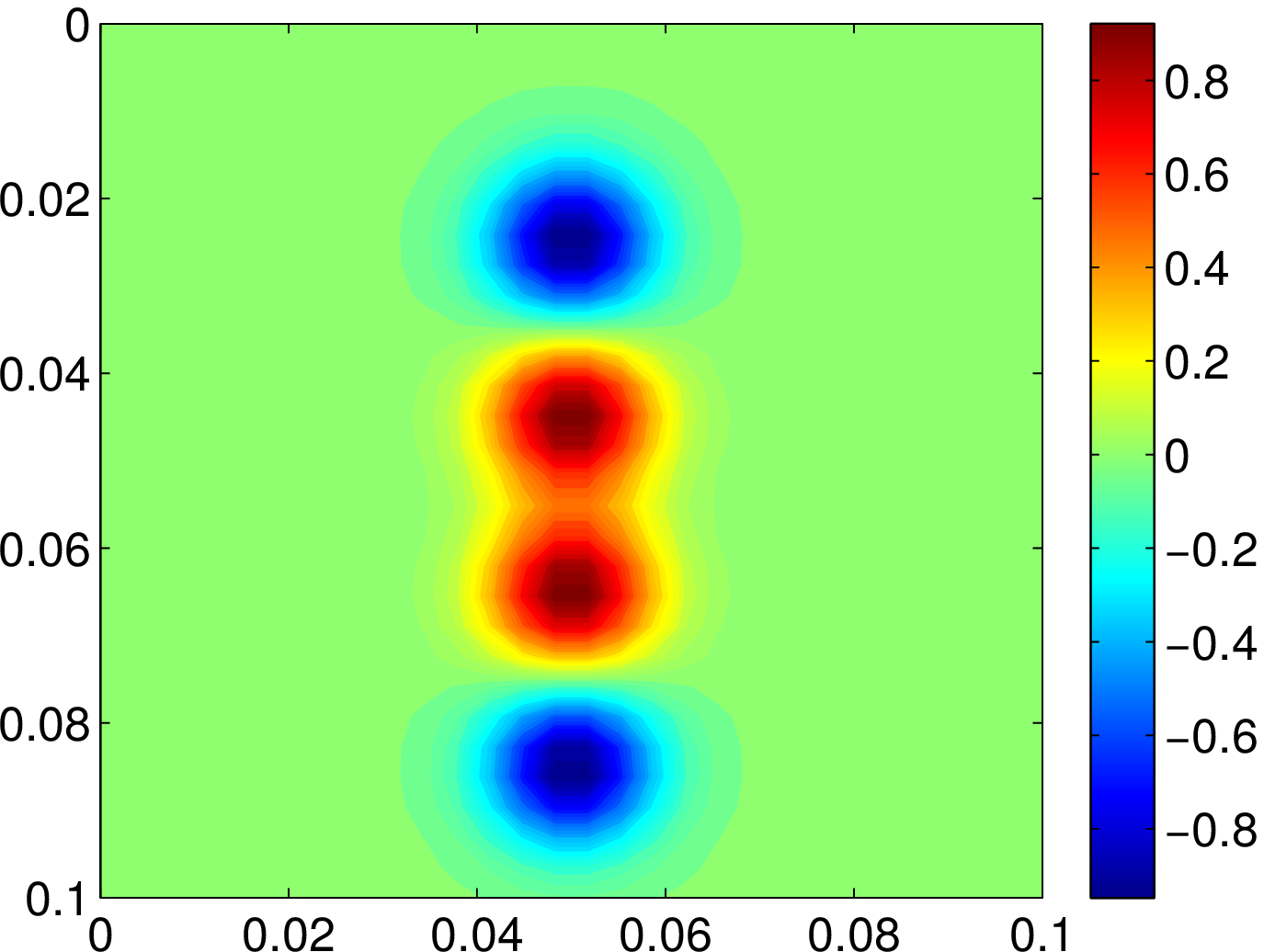}\label{fig:ip_source4}} 
 	\\
 	\subfigure[]{\includegraphics[width=0.33\columnwidth]{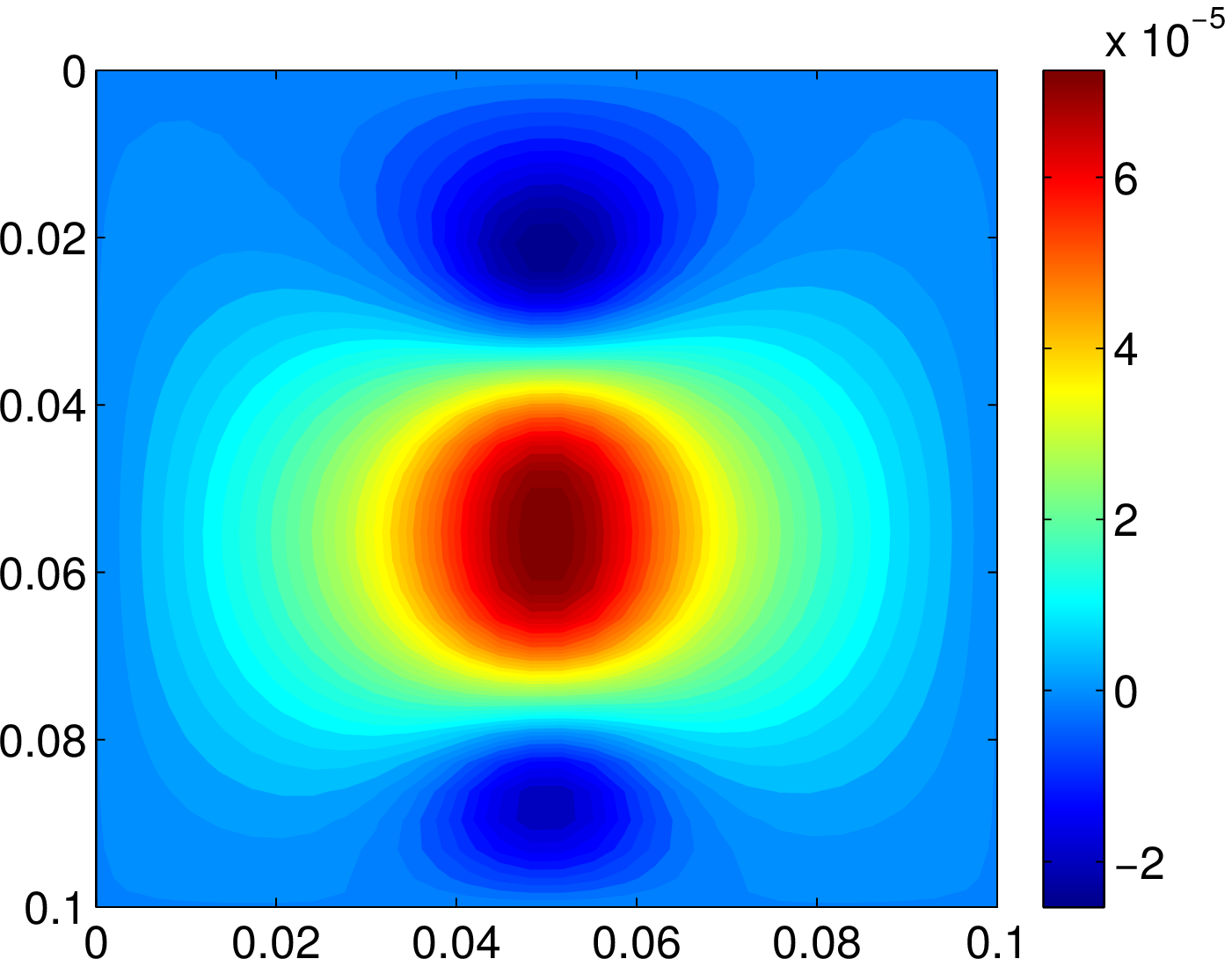}\label{fig:ip_potential4}}
 	&
 	\subfigure[]{\includegraphics[width=0.33\columnwidth]{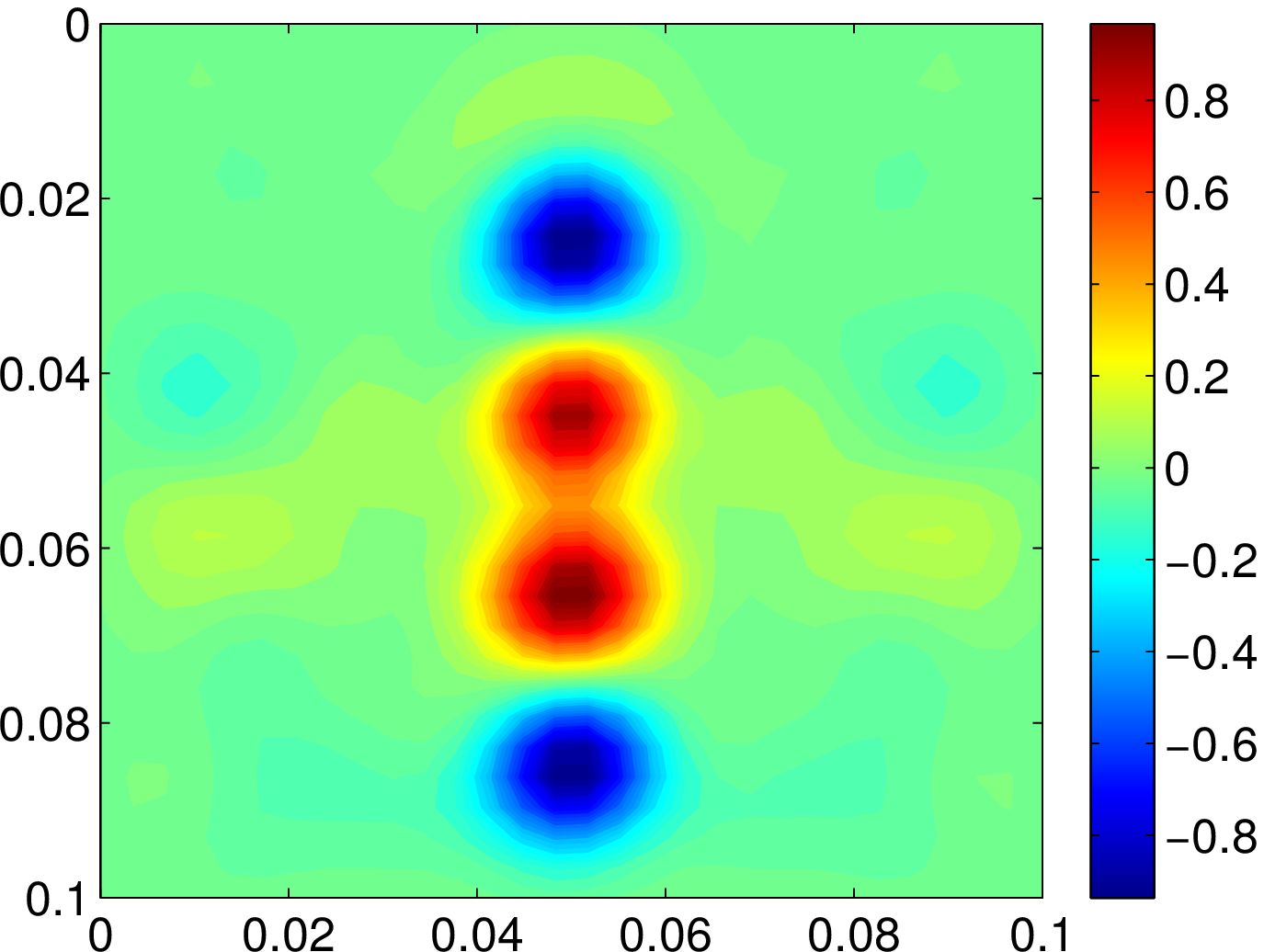}\label{fig:ip_LFMsource4}} 
	 \end{tabular}
 	\caption{ (a) electrode configuration, (b) its corresponding source of excitation function. (c) electric potential produced. (d) the source recovered using the LFM, with MSE  $= 2.5\times 10^{-3}$.}
 	\label{fig:InvProbResults4}
 \end{figure}

 The electrode configurations used for each inverse problem simulation are showed in Fig. \ref{fig:InvProbLead1}, Fig. \ref{fig:InvProbLead2}, Fig. \ref{fig:InvProbLead3}, and Fig. \ref{fig:InvProbLead4}. The source of excitation functions  prescribed for each electrode configuration are presented in Fig. \ref{fig:ip_source1}, Fig. \ref{fig:ip_source2}, Fig. \ref{fig:ip_source3}, and Fig. \ref{fig:ip_source4}. We used FEM for computing the corresponding electric potential produced by each electrode configuration, see Fig. \ref{fig:ip_potential1}, Fig. \ref{fig:ip_potential2}, Fig. \ref{fig:ip_potential3}, and Fig. \ref{fig:ip_potential4}. Finally, the posterior mean over the recovered source for each case corresponds to Fig. \ref{fig:ip_LFMsource1}, Fig. \ref{fig:ip_LFMsource2}, Fig. \ref{fig:ip_LFMsource3}, and Fig. \ref{fig:ip_LFMsource4}.  %The variance over the recovered source computed using LFM is presented in Fig. \ref{fig:varInvprob}.
 For each electrode configuration the mean squared error (MSE) between the actual source and the recovered source using the LFM was calculated.
 Fig. \ref{fig:InvProbResults1}-\ref{fig:InvProbResults4}, depict that for the four different electrode configurations, the proposed wave LFM was able to recover the source of excitation function. Therefore, one advantage of the introduced approach is that it could be used for solving the inverse problem. 


\subsection{Time-varying Source}
\begin{figure}[!ht]
 			\centering
 			\subfigure[]{\includegraphics[width=0.5\columnwidth]{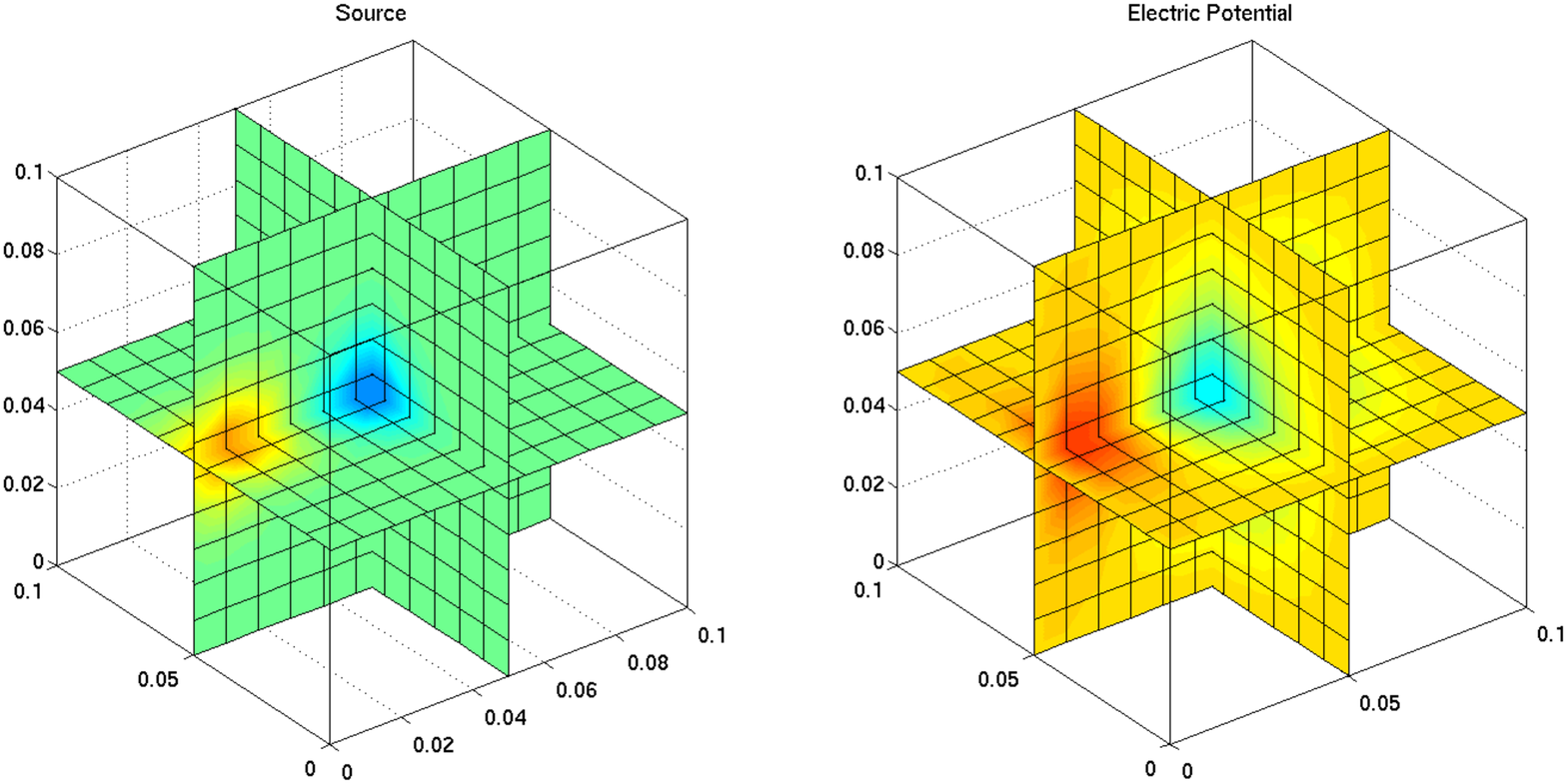}\label{fig:frame1}} 
 			\\
 			%	\subfigure[]{\includegraphics[width=0.6\columnwidth]{imagenes/time/frame2}\label{fig:frame2}} 
 			%\\
 				\subfigure[]{\includegraphics[width=0.5\columnwidth]{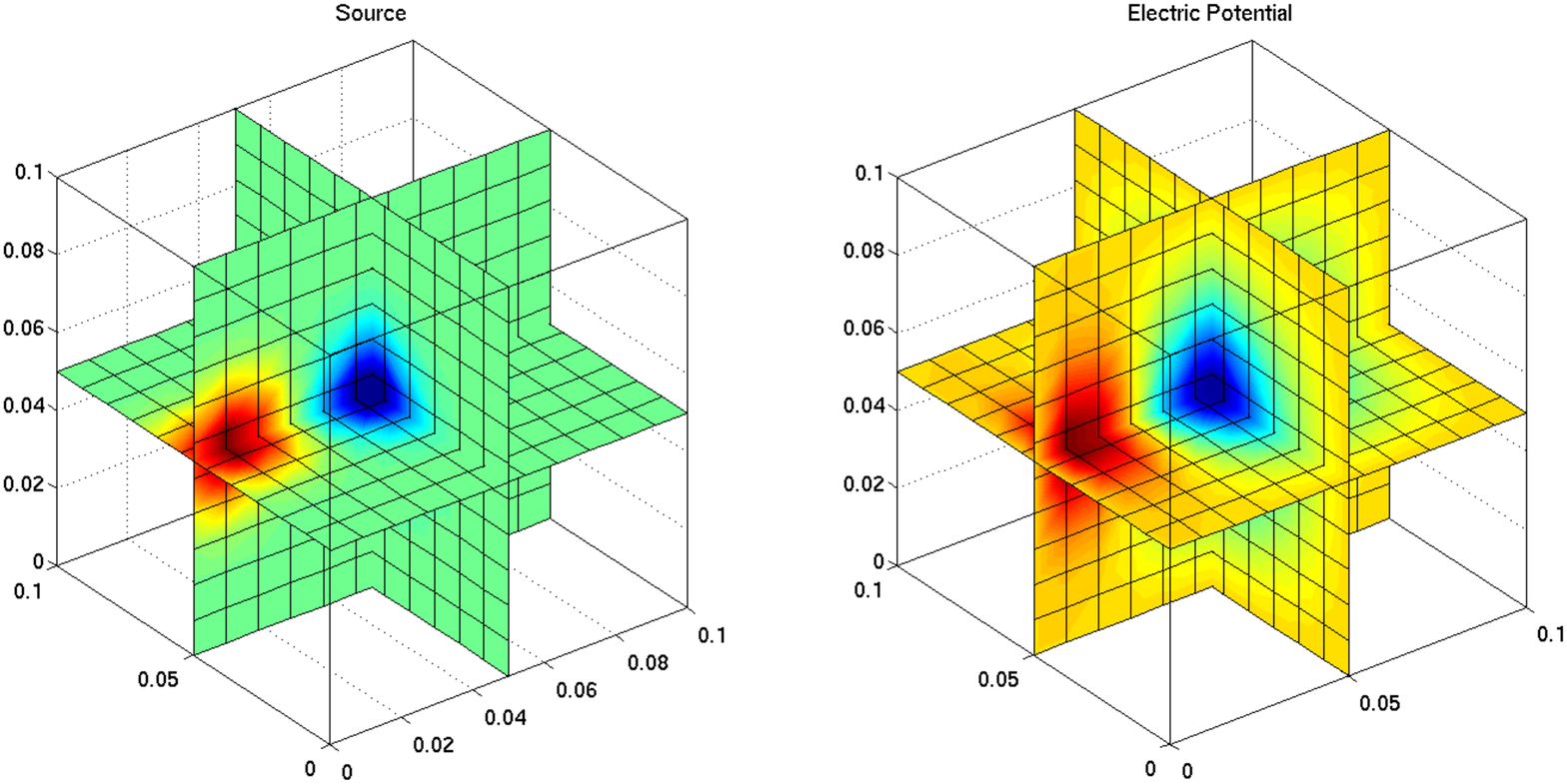}\label{fig:frame3}} 
 			\\
 			%	\subfigure[]{\includegraphics[width=0.6\columnwidth]{imagenes/time/frame4}\label{fig:frame4}} 
 		%	\\
 			%	\subfigure[]{\includegraphics[width=0.6\columnwidth]{imagenes/time/frame5}\label{fig:frame5}}  
 		%	\\
 				\subfigure[]{\includegraphics[width=0.5\columnwidth]{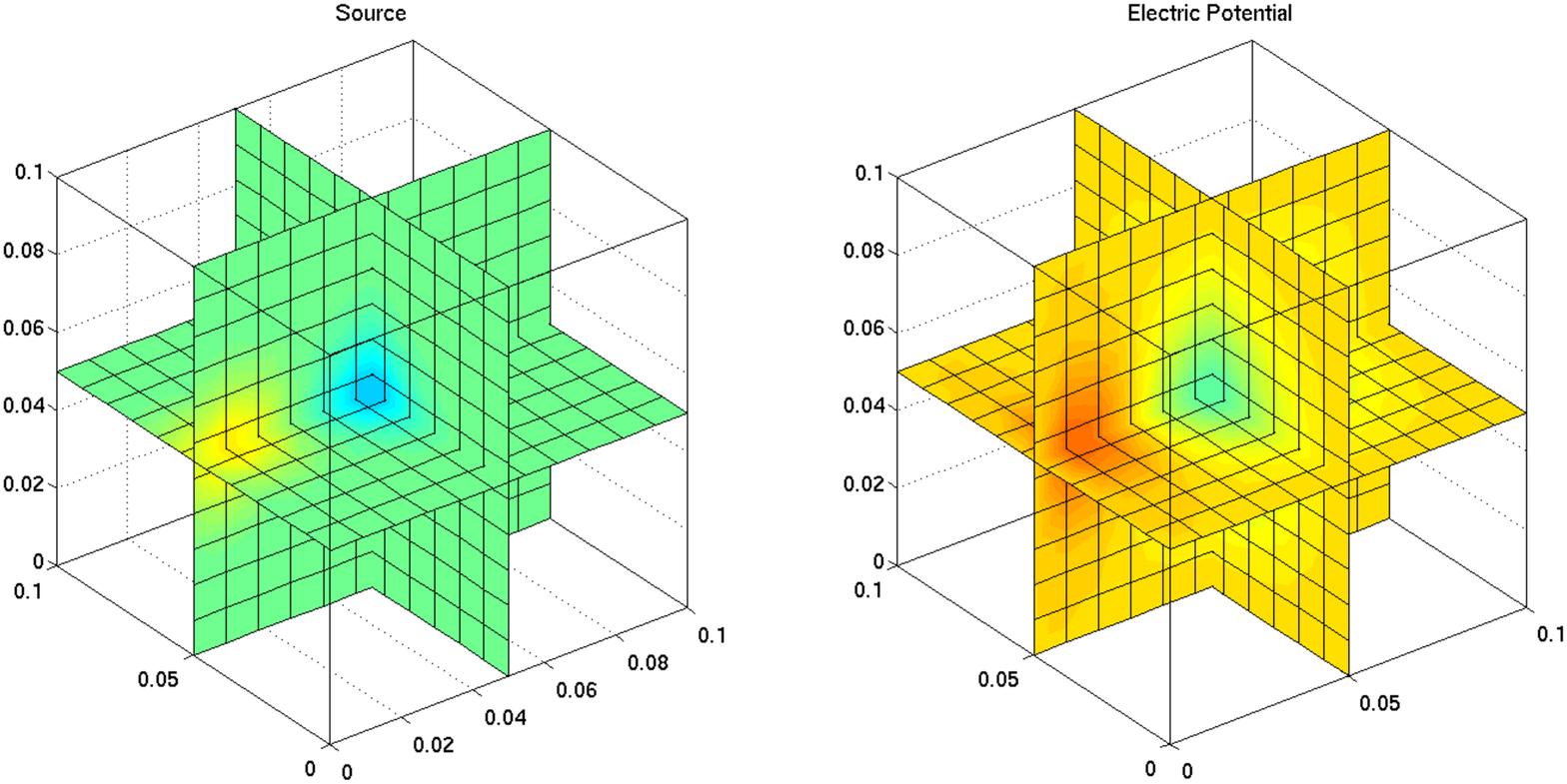}\label{fig:frame6}} 
 			\\
 			\caption{Source of excitation $u(\textbf{x},t)$  (left column) and electric potential $f(\textbf{x},t)$ (right column) for three different time instants $\textbf{t} = [0.1, \ 0.3, \ 0.6]^{\top}$.}
 			\label{fig:timeVarying}
 		\end{figure}

\noindent
To illustrate the dynamic behaviour of the wave latent force model we used here a time-varying source $u(\textbf{x},t)$  with the form 
\begin{align*}
u(\textbf{x},t) = A(\textbf{x})B(t),
\end{align*}
where $B(t) = \sin(4 \pi t /5)$ and the term $A(\textbf{x}) $ is defined as a mixture of two Gaussian distributions. 
Figures \ref{fig:frame1},\ref{fig:frame3} and \ref{fig:frame6} (left) show the source $u(\textbf{x},t)$ for three different time instants. The posterior mean over the solution to the wave equation i.e. $f(\textbf{x},t)$, for the same three time instants (Figures \ref{fig:frame1}, \ref{fig:frame3} and \ref{fig:frame6} (right)) was calculated using equation \eqref{e:f_posterior}. 
Results presented in Fig.\ref{fig:timeVarying} were obtained for the time instants $\textbf{t} = [0.1, \ 0.3, \ 0.6]^{\top}$.
From Figure \ref{fig:timeVarying} we see that with the proposed LFM the predicted electric potential field varies in time, according to the dynamic behaivor of the source of excitation. 
 %
%%-----------------------------------------------------------------------------
\subsection{Related work}
%%-----------------------------------------------------------------------------
\noindent
A wave LFM for dynamic modelling of DBS was presented in \cite{Alvarado14}. Simulations in \cite{Alvarado14} were done using oversimplified electrode configurations, where only one contact was activated. Furthermore the domain assumed in \cite{Alvarado14} had just two spatial variables ($\textbf{x} \in \mathbb{R}^{2}$), whereas the DBS phenomenon occurs in a three spatial-dimension domain. 
The approach presented in this work deals with these limitations by taking into account all the three spatial variables (except the inverse problem experiments), as well as applying the model for more complex electrode configurations. 

Although DBS simulation is typically done under the quasi-static approximation \cite{Grant09,Liberti07,McIntyre07,Schmidt12,McIntyre04}, some studies have included the time variable as part of the electric model \cite{Butson05,Bossetti08}.
To account for the electric propagation dynamics, a
Fourier Finite Element Method (Fourier FEM) was proposed
in \cite{Butson05}. The method solves Poisson
equation at different frequency components, and calculates the potential
distribution as a function of time
and space simultaneously. In \cite{Butson05} the domain of solution used was two-dimensional, homogeneous, isotropic, and the geometry was a rectangle with size ($10 \text{cm} \times 5 \text{cm}$).
Fourier FEM provides a technique to calculate time and space-dependent voltages. This is done in four steps for each
solution. First, the stimulus waveform (in this case only
a square wave) is constructed in the time domain. Next, the waveform is converted
from the time domain to the frequency domain using a
discrete Fourier transform (DFT). Third, Poisson equation is solved at
each frequency component of the DFT. The result at each component
frequency is scaled and phase shifted according to the
results of the DFT. Finally, the resulting waveform is
converted back to the time domain with an inverse Fourier
transform \cite{Butson05}.
Despite the fact that the approach implemented in \cite{Butson05} takes into account the time, Fourier FEM gives steady state solutions
and does not model transients, that is, effects of
wave propagation are neglected. 
\\
In \cite{Bossetti08} the authors compared the potentials calculated using the quasi-static approximation (Poisson equation) with those calculated from the exact solution to the inhomogeneous Helmholtz equation. Specifically, an analytical expression for the electric potential in an infinite, homogeneous, and isotropic volume conductor using a point current source stimulus was calculated from the inhomogeneous Helmholtz wave equation.  
The study presented in \cite{Bossetti08}  concludes  that the quasi-static approximation is valid, however their analysis was done for an infinite domain.  
On the other hand, the LFM we introduced in this paper has a finite domain of solution, which corresponds to the ROI in which the electric propagation is predicted.

\section{Conclusion}

\noindent 
In this work we have introduced a finite domain three-spatial dimension  latent force model based on the non-homogeneous wave equation and Gaussian process. We used our approach for describing the source of excitation as well as its corresponding electric potential produced during deep brain stimulation. We showed the benefits of our model by solving either the forward or the inverse problem.
%
%We have presented a novel latent force model for describing electric sources and fields, within the framework of deep brain stimulation, considering a finite domain of solution with three spatial dimensions. 
%
%We used the non-homogeneous partial differential wave equation and Gaussian process priors to model the electric potential as well as its source.

% Discussion static experiment 
%
In the cases where the source of excitation was assumed constant (see section \ref{DirectProblem}) results obtained using the proposed model  
proved to be close to the FEM solution of the
Poisson equation. %in section \ref{DirectProblem} (see Fig.\ref{fig:DiProb1}, Fig.\ref{fig:DiProb2}, Fig.\ref{fig:DiProb3}). 
In this sense, the more terms in the sums used for the Green's function (equation \eqref{e:Green}) in the covariance \eqref{e:kf} and cross-covariance \eqref{e:cross_cov} functions of the proposed model, the better approximation was obtained. Nevertheless, a balance between computational cost and error reduction must be done. Results show that for more than seven terms in the sums the error reduction is less significant in comparison with the increased time needed for the computation of the posterior mean and posterior variance. Additionally, Fig.\ref{fig:allDif} confirms that the contribution of the terms is smaller as the indexes in the sums present in the Green's function \eqref{e:Green} increases.    

%Discussion inverse problem
Besides, results
show that the inverse problem can be addressed using the proposed
model. The functions used for modeling the source produced by four different electrode configurations were recovered (see Fig.\ref{fig:InvProbResults1} to Fig.\ref{fig:InvProbResults4}). For the inverse problem the domain of solution was reduced to two-spatial dimensions. This was done due to the high computational cost required for calculating $K_{f}^{-1}$ in \eqref{e:u_posterior}.

The latent force model presented could be
extended in future works. First, to make use of more realistic domains, taking into account heterogeneous
and anisotropic domain properties, non-stationary LFM based on the wave equation could be studied, i.e. a model where the coefficient $a$ in \eqref{e:waveGeneral} becomes a function of the three input spatial variables. Additionally, different boundary
and initial conditions can be analyzed. Moreover, a partial differential equation
that considers the wave propagation in lossy materials might also be considered. Computational cost reduction is also an important issue that should be addressed.

\section{Acknowledgments}
\noindent
The author Pablo A. Alvarado was supported by the convening 617
“Jóvenes Investigadores e Innovadores” funded by COLCIENCIAS.
The authors are thankful to the Automática research group at
Universidad Tecnológica de Pereira. This research is developed under the
project ``Estimaci\'on de los par\'ametros de neuro modulaci\'on con
terapia de estimulaci\'on cerebral profunda en pacientes con
enfermedad de parkinson a partir del volumen de tejido activo
planeado", financed by Colciencias with code 111065740687.

\appendix \label{appendix}

\section{Solution of Covariance and Cross-covariance Functions}
%\onecolumn
\subsection*{Covariance kernel for the solution of the wave equation}
\noindent
The covariance function  $k_f(\textbf{x},\textbf{x}';t,t')$ of the output can be expressed as follows:

	\begin{align}\label{eq:kernelFinal}
	k_f(\textbf{x},\textbf{x}';t,t') =&
	\left( \frac{8 }{l_1 l_2 l_3} \right)^2
	\sum\limits_{\forall n}
	\sum\limits_{\forall m}
	\sum\limits_{\forall k}
	\sum\limits_{\forall n'}
	\sum\limits_{\forall m'}
	\sum\limits_{\forall k'}
	K_{t}(t,t')
	\times
	\cdots
	\\  \notag
	& K_{x}(x,x')
	K_{y}(y,y')
	K_{z}(z,z'),
	\end{align}
where
\begin{align}
\notag
 K_t(t,t')  &= \frac{S^{2}}{a\lambda_{nmk} a\lambda_{n'm'k'}}k_t(t,t'),
\\ \label{e:Kftime}
k_{t}(t,t') &= \int\limits_{0}^{t}  \int\limits_{0}^{t'}  \sin[a\lambda_{nmk}(t-\tau)] \sin[a\lambda_{n'm'k'}(t'-\tau')]
\times
\cdots
\\
&\exp \left[ -\frac{(\tau - \tau')^2}{\sigma_t^2} \right] \dif\tau'  \ \dif\tau .
\end{align}
$K_x(x,x')$, $K_y(y,y')$, and $K_z(z,z')$ have the general form
\begin{align}
K_l(l,l') &= C(n,m) k_{l}(l,l'), 
%\\
 %\notag
%K_x(x,x') &= C_x(n,n') k_{x}(x,x'),
%\\ \notag
%K_y(y,y') &= C_y(m,m') k_{y}(y,y'),
%\\ \notag
%K_z(z,z') &= C_z(k,k') k_{z}(z,z'),
%\begin{align}\label{eq:kernel}
%k_f(\textbf{x},\textbf{x}';t,t') = 
%\left( \frac{8 }{l_1 l_2 l_3} \right)^2
%\sum\limits_{\forall n}
%\sum\limits_{\forall m}
%\sum\limits_{\forall k}
%\sum\limits_{\forall n'}
%\sum\limits_{\forall m'}
%\sum\limits_{\forall k'}
%\frac{S S'}{a\lambda_{nmk} a\lambda_{n'm'k'}}
%k_{x}(x,x')
%k_{y}(y,y')
%k_{z}(z,z')
%k_{t}(t,t')
%C_{x}
%C_{y}
%C_{z},
%\end{align}
\\ \notag
k_{l}(l,l') &= \sin(\alpha_n l)\sin(\alpha_{m} l'),
%\\ \notag
%k_{x}(x,x') &= \sin(\alpha_nx)\sin(\alpha_{n'}x'),
%\\\notag
%k_{y}(y,y') &= \sin(\beta_m y)\sin(\beta_{m'}y'),
%\\\notag
%k_{z}(z,z') &= \sin(\gamma_k z)\sin(\gamma_{k'} z'),
% 
%
%
%
\\ \label{e:KffGeneralSpatial}
C(n,m) &= 
\int\limits_{0}^{l}  \int\limits_{0}^{l} \sin(w_n \xi) \sin(w_{m}\xi')
\exp \left[ -\frac{(\xi - \xi')^2}{\sigma^2} \right] \dif \xi'  \ \dif \xi,
%\\ \label{e:Kfx}
%C_{x}(n,n') &= \int\limits_{0}^{l_1} \int\limits_{0}^{l_1} \sin(\alpha_n \xi) \sin(\alpha_{n'} \xi')
%\exp \left[ -\frac{(\xi - \xi')^2}{\sigma_x^2} \right] \ \dif\xi' \ \dif\xi,
%%
%%
%\\ \label{e:Kfy}
%C_{y}(m,m') &= \int\limits_{0}^{l_2}  \int\limits_{0}^{l_2} \sin(q_m\eta) \sin(q_{m'}\eta')
%\exp \left[ -\frac{(\eta - \eta')^2}{\sigma_y^2} \right] \dif\eta'  \ \dif\eta,
%%
%%
%\\ \label{e:Kfz}
%C_{z}(k,k') &= \int\limits_{0}^{l_3} \int\limits_{0}^{l_3} \sin(\gamma_k \zeta) \sin(\gamma_{k'}\zeta')
%\exp \left[ -\frac{(\zeta - \zeta')^2}{\sigma_z^2} \right] \ \dif\zeta' \ \dif\zeta.
\end{align}
where $w_n$ and $w_{m}$ are constants that depend on the index $n$ and $m$, and $\sigma^2$ corresponds to the hyperparameter associated to each spatial kernel in \eqref{e:CovSource}.
%
%
%
%We can express \eqref{eq:kernelFinal} as
%\begin{align}\label{eq:kernelMatrix}
%\cov [ f( \boldsymbol{x}, \boldsymbol{y},   \boldsymbol{z},\boldsymbol{t}),f(  \boldsymbol{x}', \boldsymbol{y}',   \boldsymbol{z}',\boldsymbol{t}')  ] = 
%\left( \frac{8}{l_1l_2l_3} \right)^2
%\sum\limits_{\forall n}
%\sum\limits_{\forall m}
%\sum\limits_{\forall k}
%\sum\limits_{\forall n'}
%\sum\limits_{\forall m'}
%\sum\limits_{\forall k'}
%K_{t}(\boldsymbol{t},\boldsymbol{t}')   \otimes
%K_{x}(\boldsymbol{x},\boldsymbol{x}') \otimes
%K_{y}(\boldsymbol{y},\boldsymbol{y}') \otimes
%K_{z}(\boldsymbol{z},\boldsymbol{z}'),
%\end{align}
%where $\boldsymbol{x} = \{x_1, x_2, x_3 ...,  x_N \}$, $\boldsymbol{y} = \{y_1, y_2, y_3 ...,  x_M\}$, $\boldsymbol{z} = \{z_1, z_2, z_3 ...,  z_K\}$ and $\boldsymbol{t} = \{t_1, t_2, t_3 ...,  x_T\}$.
%
%
%
%\subsubsection{Solving $C_{x}(n,n')$, $C_{y}(m,m')$, $C_{z}(k,k')$}
%
%The expressions \eqref{e:Kfx},  \eqref{e:Kfy},  \eqref{e:Kfz} have the general form
%
The solution to the double integral in \eqref{e:KffGeneralSpatial} is defined as \cite{AlvaradoThesis14}:

%\begin{align}\label{e:Cgen}
%C(n,m) = 
%\int\limits_{0}^{l}  \int\limits_{0}^{l} \sin(w_n \xi) \sin(w_{m}\xi')
%\exp \left[ -\frac{(\xi - \xi')^2}{\sigma^2} \right] \dif \xi'  \ \dif \xi,
%\end{align}
%
%where $w_n$ and $w_{m}$ are constants that depend on the index $n$ and $m$ . The solution of \eqref{e:Cgen} is given by:
if $n\neq m$ and, $n$ and $m$ are both even or both odd, then
\begin{align}\label{eq:Cyqys_nm}
C(n,m)&=
\left(\frac{\sigma
	l}{\sqrt{\pi}(m^2-n^2)}\right)
	\times
	\cdots
	\\
	&\left\{ne^{\left(\frac{\gamma_m\sigma}{2}\right)^2}\imag\left[\mathcal{H}
(\gamma_m,l)\right]-\;me^{\left(\frac{\gamma_n\sigma}{2}\right)^2}\imag\left[\mathcal{H}(\gamma_n,l)\right]\right\},
\end{align}
otherwise
\begin{align} %\label{eq:Cyqys_nm}.
C(n,m) =0,
\end{align}
where, 
\begin{align}\label{e:Hfun}
\mathcal{H}(\zeta,\upsilon,\varphi)=\erf\left(\frac{\varphi-\upsilon}{\sigma}-\frac{\sigma \zeta}{2}\right)+\erf\left(\frac{\upsilon}{\sigma}+\frac{\sigma \zeta}{2}\right),
\end{align}
%
%
%\begin{align}\label{eq:Cyqys_nm}
%C(n,m)&=
%\begin{cases}
%\begin{aligned}
%\left(\frac{\sigma
%	l}{\sqrt{\pi}(m^2-n^2)}\right)\left\{ne^{\left(\frac{\gamma_m\sigma}{2}\right)^2}\imag\left[\mathcal{H}
%(\gamma_m,l)\right]-\;me^{\left(\frac{\gamma_n\sigma}{2}\right)^2}\imag\left[\mathcal{H}(\gamma_n,l)\right]\right\}
%\end{aligned}  & \mbox{if }n\mbox{ and }m \mbox{ are both}\\
%& \mbox{ even or both odd.}
%\\
%0&\mbox{otherwise.}
%\end{cases}
%\end{align}
when $\upsilon=\varphi$, we write $\mathcal{H}(\zeta,\upsilon)$ instead of $\mathcal{H}(\zeta,\upsilon,\upsilon)$ to keep a neat notation. For a formal calculation of \eqref{e:Hfun} see \cite{AlvaradoThesis14}. Here,  $\sigma$ also corresponds to the hyperparameter associated to each spatial kernel in \eqref{e:CovSource}.
If $n=m$, then the solution of \eqref{e:KffGeneralSpatial} corresponds to the expression
\begin{align*}
C(n)
&=\frac{\sigma\sqrt{\pi}\,l}{2}e^{\left(\frac{\gamma_n\sigma}{2}\right)^2}
\times \cdots
\\
&\left\{\real\left[\mathcal{H}(\gamma_n,l)\right]
-\imag \left[\mathcal{H}(\gamma_n,l)\right]\left[\frac{\sigma^2n\pi}{2l^2}+\frac{1}{n\pi}\right]\right\}
\times \cdots
\\
&+\frac{\sigma^2_x}{2}\left[e^{-(\frac{l}{\sigma})^2}\cos(n\pi)-1\right].
\end{align*}

\subsection*{Solving $k_{t}(t,t')$}
\noindent
In this section we present the solution of expression \eqref{e:Kftime}. The solution of $k_{t}(t,t')$ depends on whether $\lambda_{nmk}$ and  $\lambda_{n'm'k'}$ are equal or not.  
The solution of the time kernel  $k_{t}(t,t')$ for the wave equation is given by:

	\begin{align*}
	k_{t}(t,t') =& 
	c
	\cdot \real
	\left[
	\widehat{h}(\gamma',\tilde{\gamma},t,t')  
	+ 
	\widehat{h}(\gamma,\tilde{\gamma}',t',t) 
	\right. 
	\\
	& \left. 
	-\widehat{h}(\gamma',\gamma,t,t')    
	- \widehat{h}(\gamma,\gamma',t',t)
	\right] 
	\mbox{if } \lambda_{nmk} \neq \lambda_{n'm'k'},
	\end{align*}
	or,
	\begin{align*}
	k_{t}(t,t') = 
	c
	 \cdot \real
	\left[
	\widehat{H}_{2}(\gamma,-\gamma,t',t)
	-
	\widehat{h}(\gamma,\gamma,t,t')
	\right. 
	\\
	\left. 
	- \widehat{h}(\gamma,\gamma,t',t)
	\right] 
	 \mbox{if }  \lambda_{nmk} = \lambda_{n'm'k'}
	\end{align*}
	where,
	\begin{align*}
	c =& \frac{\sigma\sqrt{\pi}}{4}, 
	\\
	\widehat{h}(\zeta,\rho,\upsilon,\varphi) 
	=& 
	\frac{1}{\zeta+\rho}
	\left[
	\Upsilon(\zeta,\upsilon,\varphi)
	-e^{ - \rho \upsilon}
	\Upsilon(\zeta,0,\varphi)
	\right],
	\\
	\widehat{H}_{2}(\zeta,-\zeta,u,v) 
	=&
	\left( v + \frac{\sigma^2 \zeta}{2} \right)  \Upsilon(\zeta,v,u) 
	+
	u \Upsilon(-\zeta,u,v)
	- \cdots
	\\
	& \frac{\sigma^2 \zeta}{2}  e^{\zeta v} \Upsilon(\zeta,0,u) + \cdots
	%\right. 
	%\\ \notag &
	%\left.
	\\
	& \frac{\sigma e^{\zeta(v-u)} }{\sqrt{\pi}}
	\left[
	\hat{\mathcal{G}}(\zeta,v,u) 
	- 
	\hat{\mathcal{G}}(\zeta,0,u)
	\right],
	\\
	\Upsilon(\zeta,\upsilon,\varphi)
	=& 
	e^{(\upsilon-\varphi)\zeta} e^{\left(\frac{\zeta\sigma}{2}\right)^2}
	\mathcal{H}(\zeta,\upsilon,\varphi),
	\\
	\hat{\mathcal{G}}(\zeta,\upsilon,\varphi)
	=& 
	\exp
	\left\{
	-\left[ 
	\left(\frac{\upsilon}{\sigma}\right)^2 + \upsilon \zeta
	\right]
	\right\}
	- \cdots
	\\
	&\exp
	\left\{
	-\left[
	\left(\frac{\upsilon-\varphi}{\sigma}\right)^2
	+
	(\upsilon - \varphi)\zeta
	\right]
	\right\},
	\\
	\mathcal{H}(\zeta, \upsilon,\varphi)
	=&
	\erf\left(\frac{\varphi-\upsilon}{\sigma}-\frac{\sigma \zeta}{2}\right)+\erf
	\left(\frac{\upsilon}{\sigma}+\frac{\sigma \zeta}{2}\right).
	\end{align*}

For a detailed explanation about how to solve  \eqref{e:Kftime} see  \cite{AlvaradoThesis14}.

\subsection*{Cross covariance kernel between the latent function and the solution of the wave equation}
\noindent
The cross covariance function $k_{fu}(\textbf{x},\textbf{x}';t,t')$ between the output $f(\textbf{x},t)$ and the latent function $u(\textbf{x}',t')$, needed for the computation of the matrix $K_{fu}$ in \eqref{e:jointDis},
is given by
%\begin{align}
%k_{fu}(\textbf{x},\textbf{x}';t,t') = 
%\cov\left[f(\textbf{x},t),u(\textbf{x}',t')\right]=
%S
%\ex \left[
%\int\limits_{0}^{t}
%\int\limits_{0}^{l_3}
%\int\limits_{0}^{l_2}
%\int\limits_{0}^{l_1}
%u(\xi, \eta, \zeta, \tau)
%u(\textbf{x}',t')
%G(\textbf{x}, \xi,\eta,\zeta, t- \tau)  \dif \xi  \  \dif \eta \ \dif \zeta \ \dif \tau 
%\right]
%\end{align}
%Then, the covariance $\cov\left[f(\textbf{x},t),u(\textbf{x}',t')\right]$ is given as
\begin{align*} 
\int\limits_{0}^{t}
\int\limits_{\boldsymbol{\rho}}
%\int\limits_{0}^{l_3}
%\int\limits_{0}^{l_2}
%\int\limits_{0}^{l_1}
G(\textbf{x},\boldsymbol{\rho}, t- \tau)
\ex \left[
u(\boldsymbol{\rho}, \tau) u(\textbf{x}',t')   
\right]
\dif \boldsymbol{\rho} \
%\dif \xi  \ \dif \eta  \ \dif \zeta \
 \dif \tau,
\end{align*}

where $\boldsymbol{\rho} = [\xi,\eta,\zeta]$. Using the factorized form for the covariance of the latent function \eqref{e:CovSource}, the last expression can be written as
\begin{align*}
%S 
\int\limits_{0}^{t}
\int\limits_{\boldsymbol{\rho}}
%\int\limits_{0}^{l_3}
%\int\limits_{0}^{l_2}
%\int\limits_{0}^{l_1}
G(\textbf{x},\boldsymbol{\rho},t- \tau)
k(\xi,x') k(\eta,y') k(\zeta,z') k(\tau,t')
% \dif \xi \ \dif \eta  \ \dif \zeta \ 
\dif \boldsymbol{\rho} \
\dif \tau.
\end{align*}
With the expression \eqref{e:Green} for $G(\textbf{x},\boldsymbol{\rho}, t- \tau)$ and squared exponential kernels \eqref{e:SquaredExp} for the
covariance of the latent function, the cross covariance function  $k_{fu}(\textbf{x},\textbf{x}';t,t')$ between the latent function and the solution of the wave equation can be expressed as follows:
\begin{align}\label{e:CrosCovfu}
k_{fu}(\textbf{x},\textbf{x}';t,t') =&
\frac{8}{l_1 l_2 l_3} 
\sum\limits_{\forall n}
\sum\limits_{\forall m}
\sum\limits_{\forall k}
K_{fu}^{(x)}(x,x',n)
K_{fu}^{(y)}(y,y',m)
\times \cdots
\\ \notag
& K_{fu}^{(z)}(z,z',k)
K_{fu}^{(t)}(t,t',n,m,k),
\end{align}
where
\begin{align}\label{eq:Kwphi}
K_{f,u}^{(t)}(t,t',n,m,k)=&
\frac{S}{a\lambda_{nmk}}
\int_0^t
\sin [a\lambda_{nmk} (t-\tau) ]
\exp\left[-\frac{\left(\tau-t'\right)^2}{\sigma^2_t}\right]\dif\tau,
%\end{align}
%
%\begin{align}
\end{align}
and $K_{f,u}^{(x)}(x,x',n)$, $K_{f,u}^{(y)}(y,y',m)$, $K_{f,u}^{(z)}(z,z',k)$ have the general form
\begin{align}\label{eq:Cx2}
K_{f,u}^{(x)}(x,x',n) =&
\sin(\alpha_n x)
\int_0^{l_1}
\sin\left(\alpha_n \xi \right) \exp\left[-\frac{\left(\xi-x'\right)^2}{\sigma^2_x}\right] \dif \xi.
%\end{align}
%
%\begin{align}
%\\ \label{eq:Cy2}
%K_{f,u}^{(y)}(y,y',m) =&
%\sin(\beta_m y)
%	\int_0^{l_2}
%\sin\left(\beta_m \eta \right) \exp\left[-\frac{\left(\eta-y'\right)^2}{\sigma^2_y}\right]\dif\eta
%%\end{align}
%%
%%\begin{align}
%\\ \label{eq:Cz2}
%K_{f,u}^{(z)}(z,z',k)=&
%\sin(\gamma_k z)
%\int_0^{l_3}
%\sin\left(\gamma_k \zeta \right) \exp\left[-\frac{\left(\zeta-z'\right)^2}{\sigma^2_z}\right]\dif\zeta
\end{align}
For a description for this calculation, read  \cite{AlvaradoThesis14}.

\subsubsection*{Solving $K_{f,u}^{(t)}(t,t',n,m,k)$ and $K_{f,u,n}^{(x)}(x,x')$}
\noindent
The integrals in expressions \eqref{eq:Kwphi} and \eqref{eq:Cx2} %, \eqref{eq:Cy2} and  \eqref{eq:Cz2} 
have the general form
\begin{align}\label{eq:GenForm}
\int\limits_{0}^{u} \sin(a z + b ) \exp \left[ - \frac{(z - \phi)^2}{\sigma^2}   \right] \dif z.
\end{align}
We can express the solution to \eqref{eq:GenForm} as

\begin{align}\label{eq:GenForm2}
\frac{\sigma \sqrt{\pi}}{2}
\exp \left( \frac{\alpha \sigma}{2} \right)^2
\imag
[
\exp (\alpha \phi + \beta)
\mathcal{H}(\alpha,\phi,u)
],
\end{align}
where $\mathcal{H}(\alpha,\phi,u)$ is given by \eqref{e:Hfun}. Therefore, using \eqref{eq:GenForm2},  expressions \eqref{eq:Kwphi} and \eqref{eq:Cx2} can be written as

\begin{align}
K_{f,u}^{(t)}(t,t',n,m,k)=&
\frac{S}{a\lambda_{nmk}}
\frac{\sigma_t \sqrt{\pi}}{2}
e^ {\left( \frac{ \gamma \sigma_t}{2} \right)^2}
\times \cdots
\\ \notag
& \imag
[
\exp[\gamma(t-t')]
\mathcal{H}(-\gamma,t',t)
],
%\end{align}
%
%\begin{align}
\\ 
K_{f,u}^{(x)}(x,x',n) =&
\sin(\alpha_n x)
\frac{\sigma_x \sqrt{\pi}}{2}
e^{\left( \frac{  \widehat{\gamma}_n \sigma_x}{2} \right)^2}
\times \cdots
\\ \notag
&\imag
[
\exp ( \widehat{\gamma}_n x')
\mathcal{H}( \widehat{\gamma}_n,x',l_1)
],
%\end{align}
%
%\begin{align}
\\ 
K_{f,u}^{(y)}(y,y',m) =&
\sin(\beta_m y)
\frac{\sigma_y \sqrt{\pi}}{2}
e^ {\left( \frac{  \widehat{\gamma}_m \sigma_y}{2} \right)^2}
\times \cdots
\\ \notag
&\imag
[
\exp  \widehat{\gamma}_m y')
\mathcal{H}( \widehat{\gamma}_m,y',l_2)
],
%\end{align}
%
%\begin{align}
\\ 
K_{f,u}^{(z)}(z,z',k)=&
\sin(\gamma_k z)
\frac{\sigma_z \sqrt{\pi}}{2}
e^{ \left( \frac{  \widehat{\gamma}_k \sigma_z}{2} \right)^2}
\times \cdots
\\ \notag
&\imag
[
\exp ( \widehat{\gamma}_k z')
\mathcal{H}( \widehat{\gamma}_k,z',l_3)
],
\end{align}
where $\gamma  = j a\lambda_{nmk} $, $\widehat{\gamma}_n = j \alpha_n  $, $\widehat{\gamma}_m = j \beta_m$, and $\widehat{\gamma}_k = j \gamma_k$.

For an explanation about this calculation see  \cite{AlvaradoThesis14}.

%\twocolumn

\bibliographystyle{elsarticle-num} 
\bibliography{biblio-paper-tesis}

\end{document}